\documentclass[prb, preprint, letter, superscriptaddress]{revtex4-1}
\usepackage{amssymb}
\usepackage{amsmath}
\usepackage{bbm}
\usepackage{graphicx}
\usepackage{color}

\newcommand{\Dij}{\boldsymbol D_{ij}}
\newcommand{\Dii}{\boldsymbol D_{ii}}
\newcommand{\ris}{\{\vec r_i\}}
\newcommand{\rips}{\{\vec{r'}_i\}}
\newcommand{\Lne}{\hat L_{_\textrm{NE}}}
\newcommand{\Lfp}{\hat L_{_\textrm{FP}}}
\newcommand{\rip}{\vec{r'}_i}
\newcommand{\rjp}{\vec{r'}_j}
\newcommand{\ri}{\vec{r}_i}
\newcommand{\rj}{\vec{r}_j}
\newcommand{\hrij}{\hat r_{ij}}

\newcommand{\gi}{\nabla_i}
\newcommand{\gj}{\nabla_j}
\newcommand{\bieff}{\boldsymbol\beta^i_\textrm{eff}}
\newcommand{\bijeff}{\boldsymbol\beta^{ij}_\textrm{eff}}
\newcommand{\muii}{\boldsymbol\mu_{ii}}
\newcommand{\Umod}{U_\textrm{mod}}

\begin{document}

\author{Shenshen Wang}
\author{Peter G. Wolynes}
\affiliation{Department of Physics, Department of Chemistry and Biochemistry,  and Center for Theoretical Biological Physics, University of California, San Diego, La Jolla, CA 92093, USA}
\affiliation{Department of Chemistry, and Center for Theoretical Biological Physics, Rice University, Houston, TX 77005, USA}

\title{Tensegrity and Motor-Driven Effective Interactions in a Model Cytoskeleton}
\date{\today}

\begin{abstract}

Actomyosin networks are major structural components of the cell. They provide mechanical integrity and allow dynamic remodeling of eukaryotic cells, self-organizing into the diverse patterns essential for development. We provide a theoretical framework to investigate the intricate interplay between local force generation, network connectivity and collective action of molecular motors. This framework is capable of accommodating both regular and heterogeneous pattern formation, arrested coarsening and macroscopic contraction in a unified manner.
We model the actomyosin system as a motorized cat's cradle consisting of a crosslinked network of nonlinear elastic filaments subjected to spatially anti-correlated motor kicks acting on motorized (fibril) crosslinks. The phase diagram suggests there can be arrested phase separation which provides a natural explanation for the aggregation and coalescence of actomyosin condensates. Simulation studies confirm the theoretical picture that a nonequilibrium many-body system driven by correlated motor kicks can behave as if it were at an effective equilibrium, but with modified interactions that account for the correlation of the motor driven motions of the actively bonded nodes. Regular aster patterns are observed both in Brownian dynamics simulations at effective equilibrium and in the complete stochastic simulations. The results show that large-scale contraction requires correlated kicking.

\end{abstract}

\hyphenation{}

\maketitle

\section{Introduction}


The mechanical integrity of eukaryotic cells depends on their cytoskeleton. The cytoskeleton is made up of a dense network of protein filaments spanning the cytoplasm.
Cytoskeletal networks self-organize into highly dynamic and heterogeneous patterns from the interplay between active force generation by molecular motors and passive dissipation of energy in the crowded cellular interior. \cite{cytoskeleton}
Understanding the dynamics of such pattern formation remains a challenge to statistical mechanical theory.

Actomyosin networks are the main components of the cellular contractile machinery essential for processes as diverse as cytokinesis and wound healing. Walking on the structural scaffold provided by an actin network, myosin-II motors themselves self-assemble into bipolar minifilaments that generate sustained sliding of neighboring actin filaments relative to each other. By carrying out this correlated motion the minifilaments reorganize the filamentous actin networks and generate tension ultimately powered by ATP hydrolysis. The formation and coalescence of actomyosin aggregates to exert contractile forces are manifested in pulsed contractions of an actomyosin network that drive epithelial sheet deformation during morphogenesis.\cite{contractile ratchet 1, contractile ratchet 2} Such aggregates also are responsible for a multistage coarsening process that occurs in a bottom-up model system for contractility which has been reconstituted in vitro.\cite{multistage coarsening}

In reconstituted filament-motor assemblies, relatively \emph{regular} patterns such as asters, in which stiff filaments or filament bundles radiate from a common center, \cite{MT self-organization, active patterning} arise that resemble the mitotic spindles formed in dividing cells. A polarity sorting mechanism \cite{MT self-organization} has been proposed to explain the observed pattern. On the other hand, irregular \emph{heterogeneous} cluster structures have been seen both in the actomyosin networks of \textit{C. elegans} embryos \cite{asymmetrical contraction} and in the minimal in vitro network model. \cite{steady clusters} In the in vitro system, clusters of various sizes continuously change via fusion and rupture events but the \textmd{distribution} of cluster sizes remains steady. Motor-driven filament sliding and a well-tuned connectivity seem to play a key role in this nonequilibrium steady state.

In the present work, by looking into the intricate interplay between local force generation, network connectivity and collective motor action, we seek to provide a theoretical framework that is capable of accommodating both regular and heterogeneous pattern formation, as well as arrested coarsening and large-scale contraction in a unified manner.
Our microscopic model is a motorized cat's cradle \cite{CC, JCP, Teff ppr, spontaneous motion} which consists of a crosslinked network of nonlinear elastic filaments where motors generate equal but oppositely directed kicks at motor-bonded node pairs. This model encodes two essential aspects of actomyosin self-organization: first that individual actin filaments have highly asymmetric responses to loading, resisting large tensile forces but easily buckling under compressive loads. As a consequence, sheared biopolymer networks exhibit negative normal stresses \cite{negative normal stress} comparable in magnitude to the shear stress. Secondly the model captures the fact that motor-induced node movements come in anti-correlated pairs owing to the bipolar minifilaments. This anti-correlation mimics the relative sliding of neighboring parallel filaments driven by the attached motors. Both aspects are crucial to capturing the formation of a disordered condensed state of actomyosin aggregates.

Our coarse-grained model is completely microscopic but bears some resemblance to the macroscopic approach adopted by Levine and MacKintosh that introduces force dipoles into an elastic continuum\cite{force dipole 1, force dipole 2} where motor unbinding kinetics leads to enhanced low-frequency stress fluctuations. In the microscopic model the network connectivity and motor distribution over the bonds of the network are quenched once initially assigned, so that the nonequilibrium dynamics and structures predicted by our model arise solely from the intrinsic activity of motors firmly built into the network driving correlated motions stochastically.
This assumption is in line with the fact that the \emph{in vitro} structures are irreversibly assembled because many protein factors found \emph{in vivo} that allow fast pattern renewals are left out of the reconstitution, such as disassembly of contractile structures and transience of actin crosslinking proteins.
The model highlights the key role in determining the course of structural development played by the motor susceptibility, a parameter characterizing how sensitively the motors respond to imposed forces.
The coupling between motor kinetics and the structure leads to a double-way feedback: Motor action induces structural changes of the network and thus modifies the local mechanical environment of the motors, which in turn changes the load-dependent motor response.

In the same spirit as our earlier work on the statistical mechanics of systems with uncorrelated kicks on each node \cite{Teff ppr, spontaneous motion}, we adopt a master equation description but treat small-step spatially \emph{anti-correlated} kicks in order to mimic contractile-ratchet-like \cite{contractile ratchet 1, contractile ratchet 2} incremental deformations of actomyosin networks.
We again obtain an effective Fokker-Planck equation in the small kick limit. But the correlations lead to \emph{local} effective temperature $\mathbf{T}_\textrm{eff}$ and diffusion coefficients $\mathbf{D}_\textrm{eff}$, which now both become tensors and also depend on the \emph{instantaneous} local network structure.
More interestingly, anti-correlated kicking leads to a modification of the bare interaction.
Essentially new forces come into play through the action of the motors.
This motor-induced force depends linearly on the motor activity to quadratic order in kick step size and decays in space as the inverse distance in three dimensions, resembling a logarithmically growing potential.
By treating the motor-bonded node pairs as ``functional units" and deriving a pair-level steady-state solution of the effective Fokker-Planck equation, we demonstrate that the motor-driven anti-correlated movements of actively-bonded nodes give an additional effective pair potential that exhibits a strong short-range attraction regardless of motor susceptibility. The original interaction is also still present but is an enhanced or weakened long-range attraction. For motors with negative susceptibility at sufficiently high activity, the dominant interaction is actually a long-range repulsion. The relative contributions of the various terms yield a diverse range of steady-state structures.
This decoupling scheme in the mean-field spirit also allows us to perform a self-consistent calculation to evaluate quasi-thermodynamic phase diagrams. A non-monotonic dependence of the pressure (or tension) upon the node density (under susceptible motor kicks) indicates the possibility of phase separation.

To test the validity of the analytical approach, we compare the steady-state structural features found by Brownian dynamics simulations using the effective temperature and modified potential (both a tensor-parameter formalism and a pair-level scalar-parameter formalism) obtained from the steady-state solutions, with the structural features observed in a dynamic Monte Carlo simulation that is fully consistent with the master equation at thermal temperature and with bare interactions. We find good quantitative agreement suggesting that a non-equilibrium system driven by small-step correlated motor kicks can be thought of as being at an effective equilibrium with modified interactions.

The existence of an effective short-range attraction combined with the predicted tendency for phase separation suggests that the formation of steady heterogeneous cluster structures is an example of arrested phase separation.\cite{colloidal gels, arrested phase separation}
A force-percolating network consisting of nonlinear elastic fibers attains rigidity when local collapse induced by the motor-driven short-range attraction balances the concomitant neighboring bond stretching. This initially homogeneous network then develops into dense clumps connected by highly stretched bonds, and simultaneously, compact aggregates phase separate from node-poor regions. The coarsening process stops once a global balance is achieved, and the pertinent dynamic process involves phase separation followed by arrest due to bond constraints.
Our model naturally explains the formation of aster patterns through the notion of an effective repulsion.


\section{Theory}

\begin{figure}[htb]
\begin{center}
\centerline{\includegraphics[angle=0, scale=0.5]{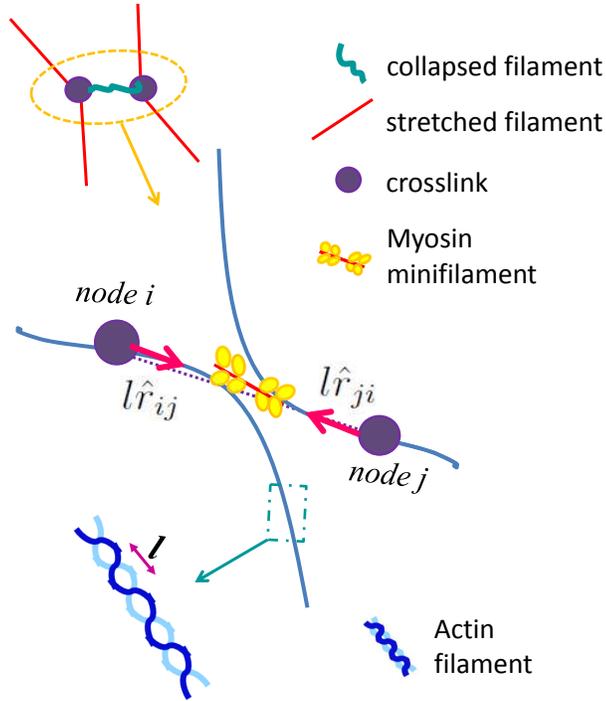}}
\caption{Schematic illustration of the spatially anti-correlated kicks acting on motor-bonded node pairs.
(Central image) A bipolar myosin minifilament pulls in slack locally, generating a pair of equal size ($l$) but oppositely directed displacements (red arrows) at the motor-bonded nodes (purple spheres) along their line of centers, where $\hat{r}_{ij}$ is a unit vector pointing from node $i$ to node $j$. Upon zoom-out, this represents a typical functional unit (marked by a dashed circle in the top image) that generates incremental contractions within a crosslinked filamentous network. An enlarged view of the actin filament (bottom image) reveals its segmented structure. The size $l$ of the subunits determines the magnitude of the relative node displacements due to contraction events of myosin sliding. $l$ is thus taken to be the step size of anti-correlated kicks in our model.}
\label{anti_correlated_kick}
\end{center}
\end{figure}

In our earlier work \cite{Teff ppr} we showed how an effective temperature describes steady-state fluctuations and responses of a model cytoskeleton, treated as an amorphous network of crosslinked nonlinear-elastic filaments, driven by \emph{uncorrelated} motor kicking events.
Here we consider \emph{anti-correlated} kicks. As sketched in Fig.~\ref{anti_correlated_kick}, each motor (myosin minifilament) generates a pair of equal but oppositely directed displacements (red arrows) at the motor-connected crosslinks/nodes (purple spheres). These anti-correlated kicks mimic the contractile-rachet-like incremental movements due to myosin-driven relative sliding of neighboring actin filaments.\cite{contractile ratchet 1, contractile ratchet 2}
We point out that myosin motors do not explicitly enter our model; instead, they are exemplified only through the anti-correlated kicks. The cartoon in Fig.~\ref{anti_correlated_kick} illustrates how these kicks are generated: A myosin minifilament attaches to two otherwise unconnected actin filaments and pulls the node on either filament toward each other.
We then assume, for simplicity, that an implicitly-motor-attached filament/bond connects the node pair, neglecting the detailed architecture of the motor-filament composite (as seen in the zoom-out view in Fig.~\ref{anti_correlated_kick} top image).
Anti-correlated kicks then act on the nodes at the two ends of such a (motor-attached) active bond.

\begin{figure}[htb]
\begin{center}
\centerline{\includegraphics[angle=0, scale=0.45]{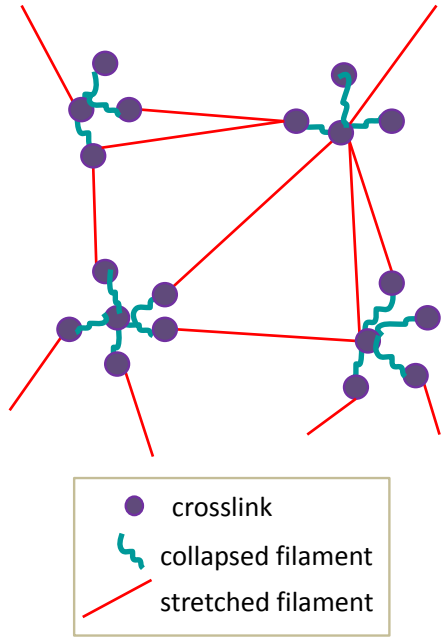}}
\caption{Cartoon of the tensegrity structure composed of collapsed and stretched elements. In a crosslinked network of filaments, active filament sliding is stabilized by passive crosslinking. A tensegrity structure is formed once a global balance between local contraction and neighboring bond stretching is achieved. An initially homogeneous network then develops into dense floppy clumps (concentrated green wiggly lines) connected by highly stretched filaments (long red straight lines).}
\label{tensegrity}
\end{center}
\end{figure}

The asymmetric load response of individual actin filaments is encoded via a nonlinear-elastic interaction between the bonded nodes, defined by the pair interaction potential $\beta U(r)=\Theta(r-L_e)\beta\gamma(r-L_e)^2/2$. Here $\Theta(\cdot)$ is the Heaviside step function and $\beta
\gamma$ gives the effective stretching stiffness of the filaments with $\beta=1/k_BT$.
An energy cost arises only when the contour length $r$ of a bond exceeds its relaxed length $L_e$. We call this interaction a ``cat's cradle" interaction.\cite{CC,JCP,Teff ppr,spontaneous motion}
For simplicity, we assume that all the bonds, no matter motor-attached or not, have the same relaxed length.
Since the motor-driven forces exceed by far the piconewton-threshold of affordable compressive loads, the induced buckling within a percolating actin network gives rise to a tensegrity structure composed of collapsed and stretched elements (illustrated in Fig.~\ref{tensegrity}): In a permanently crosslinked network of filaments, such as the in vitro reconstituted networks, active sliding of filaments is constrained by passive crosslinking, in other words, local filament or bundle contraction is balanced by the stretching of neighboring filaments. An initially homogeneous network typically then develops into dense floppy clumps (concentrated short green wiggly lines) connected by highly stretched filaments (long red straight lines).
This phenomenon leads to the formation of disordered actomyosin condensates \cite{multistage coarsening} and can lead to active contractility. \cite{active contractility}

\subsection{Quadratic expansion of the master equation: effective equilibrium with modified potential}

\subsubsection{Fokker-Planck (FP)/Smoluchowski equation for Brownian particles}

Consider a collection of $N$ Brownian particles (i.e. nodes of our model network) labeled with positional coordinates $\vec r_i$ ($i=1, \cdots, N$). For infinitesimal step Brownian motion, the configurational probability density $\Psi(\{\vec r_i\};t)$ is locally conserved and thus satisfies a continuity equation $\partial \Psi/\partial t=-\sum_i\nabla_{\vec r_i}\cdot\vec J_i$, where $\vec J_i$ is the probability current density along the coordinate of the $i$th particle.

For Brownian particles the probability current density is linearly related to the deviation of the configurational probability density from its equilibrium value
\begin{equation}
\vec J_i=-\sum_j \Dij^0(\ris)\cdot\left(\nabla_j\Psi+\beta\Psi\nabla_jU\right).
\end{equation}
Here $U(\ris)$ is the thermal equilibrium potential of mean force of the system and $\beta=(k_B T)^{-1}$. The equilibrium distribution $\Psi_\textrm{eq}(\ris)$ is related to $U(\ris)$ according to $\Psi_\textrm{eq}(\ris)\propto\exp[-\beta U(\ris)]$.
The diffusion coefficients $\Dij^0$ are functions of the system's configuration and satisfy a generalized Einstein relation with the drag coefficients $\boldsymbol\zeta_{ij}$ reading $\Dij^0=k_B T (\boldsymbol\zeta)^{-1}_{ij}$, where $\Dij^0$ and $\boldsymbol\zeta_{ij}$ are $3\times 3$ matrices for each ($i,j$) pair.


When divided into self-diffusion and coupled-diffusion parts, the FP equation $\partial\Psi/\partial t=\Lfp^0\Psi$ becomes
\begin{eqnarray}\label{Brownian dynamics}
\frac{\partial}{\partial t}\Psi(\ris;t)&=&\sum_i\nabla_i\cdot\Dii^0\cdot(\nabla_i\Psi+\beta\Psi\nabla_iU)\nonumber\\
&+&\sum_i\sum_{j\neq i}\nabla_i\cdot\Dij^0\cdot(\nabla_j\Psi+\beta\Psi\nabla_jU).
\end{eqnarray}

Note that the double gradient operation acts explicitly as
\begin{eqnarray}\label{double grad}
\nonumber\nabla_i\cdot\Dij^0\cdot\nabla_j\Psi&=&(\nabla_i\cdot\Dij^0)\cdot\nabla_j\Psi+\Dij^0:\nabla_i\nabla_j\Psi,\\
\nabla_i\cdot\Dij^0\cdot(\beta\Psi\nabla_jU)&=&(\nabla_i\cdot\Dij^0)\cdot(\beta\Psi\nabla_jU)
+\Dij^0:(\beta\nabla_i\Psi\nabla_jU+\beta\Psi\nabla_i\nabla_jU).
\end{eqnarray}

\subsubsection{Master equation for motor-driven processes: anti-correlated kicks}

To mimic the motor-driven filament sliding in actomyosin networks, we describe the motors as generating anti-correlated kicks on pairs of crosslinks that pull in slack locally (Fig.~\ref{anti_correlated_kick} middle). Since the linear size of the myosin minifilaments is small compared to the mean separation between the crosslinks, the anti-correlated moves can be treated as being along the lines of centers. In view of the segmented structure of the actin filaments, which consist of periodically arranged subunits of linear size $l$ (Fig.~\ref{anti_correlated_kick} bottom), we therefore assume a fixed kick step size $l$. $l$ indicates the amplitude of relative node displacements due to a typical contraction event. Thus an anti-correlated kick pair acting on nodes $i$ and $j$ can be represented by a pair of displacements along the line of centers $(\vec l_{ij},\vec l_{j\,i})=l(\hat r_{ij},-\hat r_{ij})$, where $\hat r_{ij}$ is a unit vector pointing from node $i$ to node $j$.
These anti-correlated kick pairs with equal size automatically satisfy momentum conservation on the macroscopic scale. Yet if we include explicitly the aqueous environment in which the cytoskeletal network is immersed, hydrodynamic interactions \cite{HD interactions} between the nodes via the solvent should be taken into account. These interactions might modify the current simplified picture, and counteract any motor-induced force imbalance on individual nodes, thus validating momentum conservation on the microscopic scale as well.

The dynamical evolution of the many-particle configuration $\ris$ due to these motor-driven events can be described by a master equation $\partial\Psi/\partial t=\Lne\Psi$ with
\begin{equation}
\Lne\Psi(\ris;t)=\int\Pi_id\vec{r'}_i\left[K(\rips\rightarrow\ris)\Psi(\rips;t)-K(\ris\rightarrow\rips)\Psi(\ris;t)\right],
\end{equation}
where the integral kernel $K(\rips\rightarrow\ris)$ encodes the probability of transitions between different node configurations. Our earlier description \cite{Teff ppr, spontaneous motion} of the motor kicking rate, $k$, still applies to current case for correlated kicks, i.e.,
\begin{equation}\label{model rate}
k=\kappa[\Theta(\Delta U)\exp(-s_u\beta\Delta U)+\Theta(-\Delta U)\exp(-s_d\beta\Delta U)],
\end{equation}
where $\kappa$ is the basal kicking rate and $s_u$($s_d$) denotes motor susceptibility to energetically uphill (downhill) moves, except that now the free energy change $\Delta U$ is due to \textit{pairs} of displacements.
Explicitly we write
\begin{eqnarray}\label{Lne0}
\nonumber&&\Lne\Psi(\ris;t)=\frac{1}{2}\kappa\sum_i\sum_j C_{ij}\int d\rip\int d\rjp\\ \nonumber
&&\times \Big\{\delta(\ri-\rip-\vec l_{ij})\delta(\rj-\rjp+\vec l_{ij})w\left[U(\cdots,\rip,\cdots,\rjp,\cdots)-U(\cdots,\ri,\cdots,\rj,\cdots)\right]\Psi(\rips;t)\\ \nonumber
&&-\delta(\ri-\rip+\vec l_{ij})\delta(\rj-\rjp-\vec l_{ij})w\left[U(\cdots,\ri,\cdots,\rj,\cdots)-U(\cdots,\rip,\cdots,\rjp,\cdots)\right]\Psi(\ris;t)
\Big\}.
\end{eqnarray}
The factor $1/2$ avoids double counting in the summation over all pairs. The quantity $C_{ij}$, much like an element of a contact map in description of protein structures, defines whether the node pair ($i,j$) is connected by an active bond and thus subject to anti-correlated displacements ($\vec l_{ij}, -\vec l_{ij}$): $C_{ij}=C_{ji}=1$ for motor-bonded pairs while $C_{ij}=C_{ji}=0$ for non-bonded pairs.
Our description of the rates gives
$w[U_i-U_f]=\Theta(U_f-U_i)\exp[-s_u\beta(U_f-U_i)]+\Theta(U_i-U_f)\exp[-s_d\beta(U_f-U_i)]$.

Assuming symmetric motor susceptibility, i.e. $s_u=s_d=s$, one finds more simply
\begin{eqnarray}\label{Lne}
\nonumber&&\Lne\Psi(\ris;t)=\frac{1}{2}\kappa\sum_i\sum_j C_{ij}\\ \nonumber
&&\times \Big\{e^{-s\beta\left[U(\ri,\rj)-U(\ri-\vec l_{ij},\rj+\vec l_{ij})\right]}
\Psi(\{\cdots,\rip=\ri-\vec l_{ij},\cdots,\rjp=\rj+\vec l_{ij},\cdots\};t) \\
&& -e^{-s\beta\left[U(\ri+\vec l_{ij},\rj-\vec l_{ij})-U(\ri,\rj)\right]}\Psi(\{\cdots,\ri,\cdots,\rj,\cdots\};t)\Big\}.
\end{eqnarray}
We assume that kicks on different pairs of nodes at any given time are uncorrelated.
The rates of possible kicking events depend on the \textit{instantaneous} node configuration reflecting an assumed Markovian character of the motor dynamics. There is no angular average due to the definiteness of kicking directions for a given configuration.
Note that the motor power strokes and thus the kick steps are discrete occurring in a stochastic fashion. The correlated motions pull in slack locally while simultaneously pulling taut neighboring filaments until a global balance is reached or a macroscopic collapse occurs, depending on whether the motors are downhill prone (with a large positive $s$) or load-resisting (with a small or negative $s$), respectively.

Quadratic expansion of Eq.~\ref{Lne} in kick step size $l$ leads to
\begin{eqnarray}
\nonumber&&\Lne\Psi(\ris;t)=\frac{1}{2}\kappa l^2\sum_i\sum_j C_{ij}\\
\nonumber&&\times\Big[\frac{1}{2}\hrij\hrij:\gi\gi\Psi+\frac{1}{2}\hrij\hrij:\gj\gj\Psi-\hrij\hrij:\gi\gj\Psi\\
\nonumber&&+s\hrij\hrij:\gi\Psi\gi\beta U+s\hrij\hrij:\gj\Psi\gj\beta U
-s\hrij\hrij:\gi\Psi\gj\beta U-s\hrij\hrij:\gj\Psi\gi\beta U\\
\nonumber&&+s\left(\hrij\hrij:\gi\gi\beta U+\hrij\hrij:\gj\gj\beta U-2\hrij\hrij:\gi\gj\beta U\right)\Psi\Big].
\end{eqnarray}
Notice that $\sum_i\sum_jC_{ij}\hat r_{ij}\hat r_{ij}:\nabla_i\nabla_i=\sum_i\sum_jC_{ij}\hat r_{ij}\hat r_{ij}:\nabla_j\nabla_j$, the above expression can be rewritten as
\begin{eqnarray}\label{NE processes}
\nonumber\Lne\Psi(\ris;t)&=&\sum_i\left[\frac{1}{2}\kappa l^2\sum_{j\neq i}C_{ij}\hrij\hrij:\gi\gi\Psi+\beta s\kappa l^2\sum_{j\neq i}C_{ij}\hrij\hrij:\left(\gi\Psi\gi U+\Psi\gi\gi U\right)\right]\\
&-&\sum_i\sum_{j\neq i}\left[\frac{1}{2}\kappa l^2 C_{ij}\hrij\hrij:\gi\gj\Psi
+\beta s\kappa l^2 C_{ij}\hrij\hrij:\left(\gi\Psi\gj U+\Psi\gi\gj
 U\right)\right].
\end{eqnarray}
The definitions in Eq.~\ref{double grad} allow us to express $\Lne\Psi$ in the form of the divergence of a flux plus some extra terms which modify the bare interactions as shown below.

\subsubsection{Generalized FP equation for motorized systems: effective temperature and modified potential}

By combining the pure Brownian dynamics (Eq.~\ref{Brownian dynamics}) with the nonequilibrium dynamics due to correlated motor-driven processes up to $O(l^2)$ (Eq.~\ref{NE processes}), we obtain an effective FP equation with effective tensor parameters ($\boldsymbol D^\textrm{eff}$ and $\boldsymbol\beta_\textrm{eff}$) and sitewise modified potentials ($U_\textrm{mod}$)
\begin{eqnarray}\label{eff FP}
\nonumber(\Lfp^0+\Lne)\Psi(\ris;t)&=&
\sum_i\Big[\gi\cdot\Dii^\textrm{eff}\cdot\gi\Psi+\gi\cdot\left(\Dii^\textrm{eff}\cdot\bieff\right)\cdot\left(\Psi\gi U^i_{mod}\right)\Big]\\
&+&\sum_i\sum_{j\neq i} \Big[\gi\cdot\Dij^\textrm{eff}\cdot\gj\Psi+\gi\cdot\left(\Dij^\textrm{eff}\cdot\bijeff\right)\cdot\left(\Psi\gj U^j_{mod}\right)\Big].
\end{eqnarray}
The effective diffusion constants read
\begin{equation}\label{Deff}
\Dii^\textrm{eff}=\Dii^0+\frac{1}{2}\kappa l^2\sum_{j\neq i}C_{ij}\hrij\hrij,\,\,
\Dij^\textrm{eff}=\Dij^0-\frac{1}{2}\kappa l^2C_{ij}\hrij\hrij.
\end{equation}
The effective temperatures are given by
\begin{eqnarray}\label{Teff}
\nonumber&&\bieff=\beta\Big(\Dii^0+s\kappa l^2\sum_{j\neq i}C_{ij}\hrij\hrij\Big)
\cdot\Big(\Dii^0+\frac{1}{2}\kappa l^2\sum_{j\neq i}C_{ij}\hrij\hrij\Big)^{-1},\\
&&\bijeff=\beta\Big(\Dij^0-s\kappa l^2C_{ij}\hrij\hrij\Big)
\cdot\Big(\Dij^0-\frac{1}{2}\kappa l^2C_{ij}\hrij\hrij\Big)^{-1}.
\end{eqnarray}
The modified potentials $U^{i(j)}_\textrm{mod}=U+\Lambda^{i(j)}$ involve the (additive) modifications $\Lambda$ that satisfy
\begin{eqnarray}\label{lambda}
\nonumber&&-\gi\Lambda^i=-\Big[\gi\cdot\big(\frac{1}{2}\kappa l^2\sum_{j\neq i}C_{ij}\hrij\hrij\big)\Big]
\cdot\Big[\big(\Dii^0+s\kappa l^2\sum_{j\neq i}C_{ij}\hrij\hrij\big)\beta\Big]^{-1},\\
&&-\gj\Lambda^j=\Big[\gi\cdot\big(\frac{1}{2}\kappa l^2C_{ij}\hrij\hrij\big)\Big]
\cdot\Big[\big(\Dij^0-s\kappa l^2C_{ij}\hrij\hrij\big)\beta\Big]^{-1}.
\end{eqnarray}

We assume that the active connectivity described by $\{C_{ij}\}$ is quenched once initially assigned and use the identity that $\gj\cdot(\hrij\hrij)=-\gi\cdot(\hrij\hrij)=\hrij(d-1)/r_{ij}$, where $d$ is the spatial dimension and $r_{ij}=|\ri-\rj|$, to obtain explicit expressions.


Several nontrivial features can be read off from the above expressions (Eqs.~\ref{eff FP}--\ref{lambda}):

(1) There is a key difference from the simple situation for uncorrelated isotropic kicks. \cite{Teff ppr, spontaneous motion} In that case an average over kicking directions yields \emph{uniform scalars} $T_\textrm{eff}$ and $D_\textrm{eff}$ which are fully determined by the motor properties (motor susceptibility $s$ and activity defined as $\Delta\equiv\kappa l^2/D_0$) regardless of the specific structure of the system. Under correlated kicks along the lines of centers, however, both $\boldsymbol T_\textrm{eff}$ and $\boldsymbol D_\textrm{eff}$ become \emph{local tensors}. These tensors depend on local network structure (relative position of bonded neighbors $\vec r_{ij}$) and motor distribution (quenched active connectivity defined by $C_{ij}$) about the central node $i$.
The tensorial nature of the effective diffusion coefficients ($\boldsymbol D_\textrm{eff}$) and mobility ($\boldsymbol\mu_\textrm{eff}\equiv\boldsymbol D_\textrm{eff}\cdot\boldsymbol\beta_\textrm{eff}$) leads to the diffusive flux not aligning with the density gradient or the drift flux.
Motor-induced modifications to the forces ($-\gi U_\textrm{mod}$) and to the transport coefficients ($\boldsymbol D_\textrm{eff}$ and $\boldsymbol\mu_\textrm{eff}$) only have longitudinal components since $\hrij\hrij$ essentially serves as a longitudinal projection operator.

(2) Anti-correlation of the myosin-generated kicks at the two ends of the actin filaments provides a microscopic basis for the anisotropy of actin diffusivity: The anti-correlated kicks enhance self-diffusion of individual molecules, whereas they slow the relative coupled diffusion of motor-bonded molecules (Eq.~\ref{Deff}).
Recent experiments \cite{active diffusion} indeed have reported that motor activity can give rise to cytoplasmic motion that has the appearance of diffusion but is significantly enhanced in its magnitude. Such ``active" cytoplasmic diffusion could enable rapid intracellular transport of matter and information.



(3) To quadratic order in kick step size $l$, it is easy to show:

(a) Both $\gi\left(\Lambda^i/k_B T\right)$ and $\gj\left(\Lambda^j/k_B T\right)$ are proportional to $\kappa l^2/D_0$. This indicates that the motor-induced forces depend linearly on the strength of the kicking noise relative to the thermal noise in small kick limit.

(b) The inverse effective temperature tensor becomes
\begin{eqnarray}
\nonumber\bieff&=&\beta\left(D_0\boldsymbol{1}+s\kappa l^2\sum_{j\neq i}C_{ij}\hrij\hrij\right)\cdot\left(\frac{1}{D_0}\right)\left(\boldsymbol{1}-\frac{1}{2}\frac{\kappa l^2}{D_0}\sum_{j\neq i}C_{ij}\hrij\hrij+O(l^4)\right)\\
\nonumber&=&\beta\left[\boldsymbol{1}+(s-\frac{1}{2})\frac{\kappa l^2}{D_0}\sum_{j\neq i}C_{ij}\hrij\hrij\right]+O(l^4)
\end{eqnarray}
Anti-correlated kick pairs give rise to effective temperature tensors that depend on local active connectivity, implying that the speed of heat flow in motor-kicking directions ($C_{ij}\neq 0$ thus $T_\textrm{eff}\neq T$) is different from energy flux along motor-free directions ($C_{ij}=0$ thus $T_\textrm{eff}=T$). Along individual active bonds, as for the uncorrelated kick case, we have $T_\textrm{eff}<T$ if $s>1/2$ whereas $T_\textrm{eff}>T$ if $s<1/2$, as well as the same detailed balance condition that if $s=1/2$ then $T_\textrm{eff}=T$.


Therefore the nonequilibrium system driven by small-step anti-correlated motor kicks may be described by an effective FP (Smoluchowski) equation at local effective temperature and with modified interaction potential.

\subsection{Pair-level steady-state solution}

We may ask whether it is still possible to obtain an explicit (but perhaps approximate) steady-state solution to an effective Fokker-Planck equation with tensor transport coefficients and modified interaction potential. Existence of such a solution makes possible the mapping of the system to an equilibrium system even for this nonequilibrium situation with spatially correlated motorized events. This mapping thus allows the study of rheological properties within a quasi-equilibrium framework. The main difficulty in making this mapping arises from the tensorial nature of the effective temperature which depends on local structures.


To approximate the $N$-body solution, we employ a decoupling scheme that reduces the problem to finding the steady states for the diffusion of ``functional units".
We will first study the simplest case for a single pair of motor-bonded nodes without hydrodynamic interactions, i.e., neglecting the $\boldsymbol D_{ij}$-related terms. A straightforward calculation leads to an explicit steady-state solution to this two-body problem.

Consider a pair of nodes located at $\vec{r}_1$ and $\vec{r}_2$. For purely Brownian motion, the $2$-body configurational probability density $\Psi(\vec{r}_1,\vec{r}_2;t)$ evolves according to the bare Fokker-Planck equation $\partial\Psi/\partial t=\Lfp^0\Psi$ with
\begin{equation}\label{FP}
\Lfp^0\Psi=\sum_{i=1,2}\left[\gi\cdot\Dii^0\cdot\gi\Psi+\gi\cdot\muii^0\cdot\gi U\Psi\right].
\end{equation}
Here the mobility tensor is related to the diffusion tensor simply by a multiplication of the inverse scalar temperature as $\muii^0=\beta\Dii^0$. The steady-state solution thus obeys the usual Boltzmann law $\Psi_{ss}^0\propto\exp[-\beta U(r_{12})]$ where the interaction potential $U$ only depends on the node separation $r_{12}=|\vec{r}_1-\vec{r}_2|$ in the absence of external fields.

The dynamic evolution of a pair due to anti-correlated motor kicks is described by a master equation $\partial\Psi/\partial t=\Lne\Psi$ with
\begin{eqnarray}\label{NE}
\nonumber\Lne\Psi(\vec{r}_1,\vec{r}_2;t)=\kappa C_{12}&\Big\{&e^{-s\beta\left[U(\vec{r}_1,\vec{r}_2)-U(\vec{r}_1-\vec{l}_{12},\vec{r}_2+\vec{l}_{12})\right]}
\Psi(\vec{r}\,'_1=\vec{r}_1-\vec{l}_{12},\vec{r}\,'_2=\vec{r}_2+\vec{l}_{12};t)\\
&&-e^{-s\beta\left[U(\vec{r}_1+\vec{l}_{12},\vec{r}_2-\vec{l}_{12})-U(\vec{r}_1,\vec{r}_2)\right]}
\Psi(\vec{r}_1,\vec{r}_2;t)\Big\}
\end{eqnarray}

By combining the pure Brownian dynamics (Eq.~\ref{FP}) and the motor-driven dynamics (Eq.~\ref{NE}) we obtain in the small kick limit an effective Fokker-Planck equation characterized by effective diffusion constants, tensor effective temperature as well as a modified potential
\begin{equation}\label{eff_FP}
(\Lfp^0+\Lne)\Psi(\vec{r}_1,\vec{r}_2;t)=\sum_{i=1,2}\left[\gi\cdot\Dii^\textrm{eff}\cdot\gi\Psi
+\gi\cdot(\Dii^\textrm{eff}\cdot\bieff)\cdot(\Psi\gi U^i_\textrm{mod})\right].
\end{equation}
In this two-body case, the effective diffusion tensors, up to $O(l^2)$, are given by
\begin{equation}\label{D_eff}
\boldsymbol D^\textrm{eff}_{11}
=D_0\boldsymbol{1}+C_{12}\frac{1}{2}\kappa l^2\hat{r}_{12}\hat{r}_{12}
=D_0\boldsymbol{1}+C_{21}\frac{1}{2}\kappa l^2\hat{r}_{21}\hat{r}_{21}
=\boldsymbol D^\textrm{eff}_{22}.
\end{equation}
The effective mobility tensors, up to $O(l^2)$, are given by
\begin{equation}\label{mu_eff}
\boldsymbol \mu^\textrm{eff}_{11}
=\beta\left(D_0\boldsymbol{1}+C_{12}s\kappa l^2\hat{r}_{12}\hat{r}_{12}\right)
=\beta\left(D_0\boldsymbol{1}+C_{21}s\kappa l^2\hat{r}_{21}\hat{r}_{21}\right)
=\boldsymbol \mu^\textrm{eff}_{22}.
\end{equation}
The tensor effective temperatures thus read
\begin{equation}\label{beta_eff}
\nonumber\boldsymbol \beta^\textrm{eff}_{1}=\boldsymbol \beta^\textrm{eff}_{2}
=\beta\left[\boldsymbol{1}+C_{12}\left(s-\frac{1}{2}\right)\frac{\kappa l^2}{D_0}\hat{r}_{12}\hat{r}_{12}\right]+O(l^4).
\end{equation}
The modifications to the bare forces are central forces that decay in space as $1/r$:
\begin{equation}
\nonumber-\nabla_1\Lambda=C_{12}\frac{\kappa l^2}{\beta D_0}\frac{1}{r_{12}}\hat{r}_{12}\, ,\quad
-\nabla_2\Lambda=C_{21}\frac{\kappa l^2}{\beta D_0}\frac{1}{r_{21}}\hat{r}_{21}=-\left[-\nabla_1\Lambda\right],
\end{equation}
suggesting the following form of the modified potential
\begin{equation}\label{U_mod}
U_\textrm{mod}(r_{12})=U(r_{12})+C_{12}\frac{\kappa l^2}{D_0}k_B T\ln(r_{12})+\textrm{const}.
\end{equation}
For $s=1/2$, the effective temperature becomes uniform scalar again $\boldsymbol \beta^\textrm{eff}=\beta\boldsymbol{1}$. The steady-state pair solution thus has precisely the form $\Psi_{ss}(r_{12})\propto\exp[-\beta U_\textrm{mod}(r_{12})]$ where $U_\textrm{mod}$ is given by Eq.~\ref{U_mod}.

For $s\neq1/2$, we have
\begin{eqnarray}
\nonumber\boldsymbol \beta^\textrm{eff}\cdot\left[-\nabla_1\Umod\right]
&=&\beta\left[\boldsymbol{1}+C_{12}\left(s-\frac{1}{2}\right)\frac{\kappa l^2}{D_0}\hat{r}_{12}\hat{r}_{12}\right]
\cdot\left(-\nabla_1 U+C_{12}\frac{\kappa l^2}{\beta D_0}\frac{1}{r_{12}}\hat{r}_{12}\right)\\
&=&\beta\left[1+C_{12}\left(s-\frac{1}{2}\right)\frac{\kappa l^2}{D_0}\right]\left[-\nabla_1 U\right]
+C_{12}\frac{\kappa l^2}{D_0}\frac{1}{r_{12}}\hat{r}_{12}+O(l^4)
\end{eqnarray}
and
\begin{equation}
\boldsymbol \beta^\textrm{eff}\cdot\left[-\nabla_2\Umod\right]
=-\boldsymbol \beta^\textrm{eff}\cdot\left[-\nabla_1\Umod\right].
\end{equation}
Here we have used the identity $\hat{r}_{12}(\hat{r}_{12}\cdot\nabla_1 U)=\nabla_1 U$, since $\nabla_1U$ is parallel to $\hat{r}_{12}$ in the two-body situation.
We see therefore that we still arrive at a steady-state solution given by
\begin{equation}\label{pair solution}
\Psi_{ss}\propto
\exp\Bigg\{-\beta\left[1+C_{12}\left(s-\frac{1}{2}\right)\frac{\kappa l^2}{D_0}\right]U-C_{12}\frac{\kappa l^2}{D_0}\ln(r_{12})\Bigg\}\equiv\exp\left[-\bar{\beta}_\textrm{eff}U_\textrm{eff}\right],
\end{equation}
where the scalar inverse effective temperature ($\bar{\beta}_\textrm{eff}$) and the effective interaction potential ($U_\textrm{eff}$) are given by
\begin{equation}\label{Teff}
\bar\beta_\textrm{eff}=\beta\left[1+C_{12}\left(s-\frac{1}{2}\right)\frac{\kappa l^2}{D_0}\right]
\end{equation}
and
\begin{equation}\label{Ueff}
U_\textrm{eff}=U+\frac{C_{12}\left(\kappa l^2/D_0\right)}{1+C_{12}(s-1/2)\left(\kappa l^2/D_0\right)}\,k_B T\ln(r_{12}).
\end{equation}


Since on the pair level the \emph{total} force is automatically along the line of centers, $\bar{T}_\textrm{eff}$ becomes effectively a scalar. Thus at the pair level the steady states take a form consistent with our earlier result for uncorrelated isotropic kicks \cite{Teff ppr} (note that here is no $1/d$ factor that arises from the angular average).
The modification to the bare interaction now, however, yields an additional central force which decays in space as $1/r$.
In the limit of high motor activity, i.e., $\Delta\equiv\kappa l^2/D_0\gg1$, Eq.~\ref{Ueff} reduces to $U_\textrm{eff}-U\simeq k_BT\ln(r_{12})/(s-1/2)$. We see the motor-induced forces can be attractive or repulsive depending on whether the motor susceptibility $s$ is larger or smaller than $1/2$, respectively.

The scaled effective pair potential given by $\bar\beta_\textrm{eff}U_\textrm{eff}=\beta U\left[1+C_{12}(s-1/2)\Delta\right]+C_{12}\Delta\ln(r_{12})$ consists of two terms. The first term indicates that the motor action may enhance or weaken the long-range attraction, arising from the ordinary bond stretching in the model cytoskeleton. The sign of this effect depends on the sign of $(s-1/2)$. The second effect of the anti-correlated motors is an effective confinement potential that promotes further contraction even in the buckling regime. This term is independent of motor susceptibility. Therefore by varying $s$ we may distinguish the influences of either term.

\begin{figure}[htb]
\begin{center}
\centerline{\includegraphics[angle=0, scale=0.55]{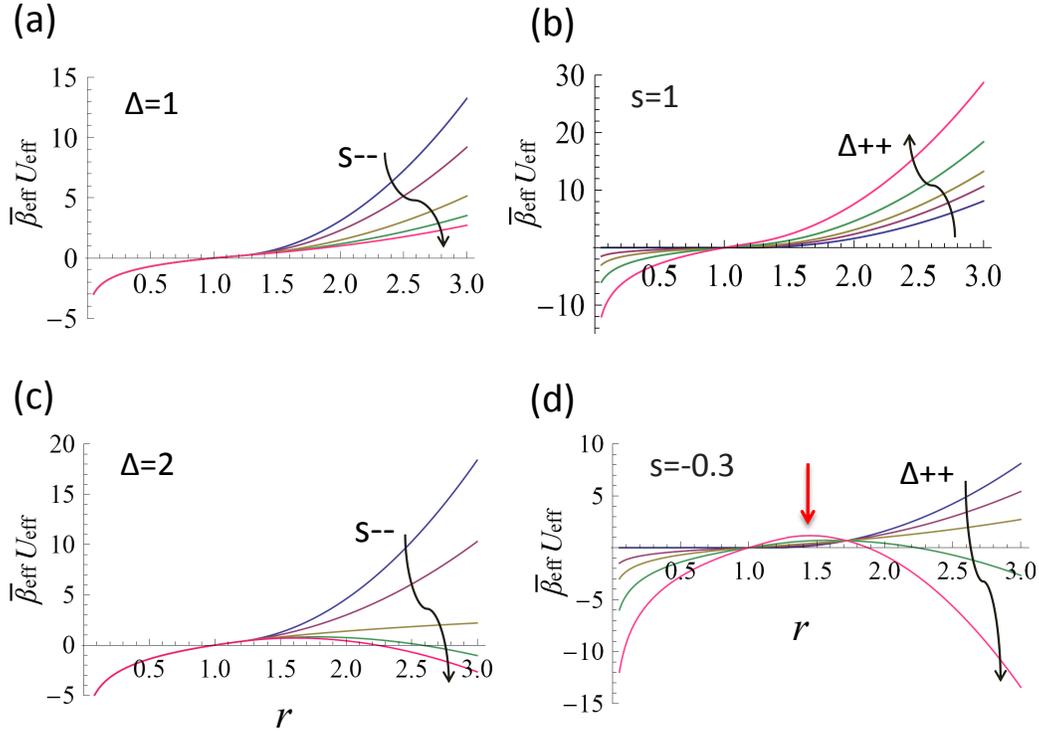}}
\caption{Profile of the modified interaction given by the pair-level steady-state solution. We plot the effective interaction $U_\textrm{eff}$ (Eq.~\ref{Teff}) scaled by effective temperature $\bar{\beta}_\textrm{eff}^{-1}$ (Eq.~\ref{Ueff}) for various motor activity ($\Delta$) and susceptibility ($s$). At sufficiently high activity, load-resisting ($s=-0.3$) motors may yield a long-range effective repulsion, an energy barrier (indicated by a red arrow in panel d) thus appears at intermediate distances, suggesting the tendency for node separation and thus the bond stretching that occurs in aster formation. (a) and (c): $s=1, 0.5, 0, -0.2$ and $-0.3$. (b) and (d): $\Delta=0, 0.5, 1, 2$ and $4$. Common parameters are $L_e=1.2, \beta\gamma=5, P_c=0.4,$ and $P_a=1$.}
\label{pair_Ueff}
\end{center}
\end{figure}

Fig.~\ref{pair_Ueff} displays the profile of the effective pair interaction $U_\textrm{eff}$ (Eq.~\ref{Ueff}) scaled by the effective temperature $\bar{\beta}_\textrm{eff}^{-1}$ (Eq.~\ref{Teff}). As shown in panels (a) and (c), when the motor susceptibility $s$ is varied the long-range interaction becomes modified but the effective attraction at small separations is not affected. In contrast, increasing the motor activity $\Delta$ not only influences interactions at large distances but also enhances the short-range attraction, as seen in panels (b) and (d). Notably, at sufficiently high activity, load-resisting motor kicks ($s<0$) may yield a long-range repulsion, i.e. the slope of the $\bar{\beta}_\textrm{eff}U_\textrm{eff}$ curve becomes negative at large $r$ (see two lowest curves in panels c and d). As a consequence, an energy barrier (indicated by the red arrow) appears at intermediate distances, indicating the tendency for node separation and thus bond stretching that ultimately leads to aster formation observed in simulations.

\subsection{Self-consistent phonon (SCP) calculation: possibility of phase separation}

The pair-level steady-state solution derived above allows us to obtain the effective pair potential and self consistently determine the Debye-Waller factor of the localized nodes using the self-consistent phonon (SCP) method. \cite{fixman}
It is straightforward to derive the effective potential associated with a ``cat's cradle" with excluded volume,\cite{JCP} i.e., a system where the nonlinear bonds (with an effective stiffness $\beta\gamma$) stretch elastically when their contour length exceeds the relaxed length $L_e$ but that buckle upon shortening too much. We assume a negligibly small hard core size ($\sigma$) for comparison with simulations.
We assign two mean-field parameters to characterize the network architecture: (1) network connectivity, $P_c$, which denotes the fraction of nearest-neighbor pairs that are bonded by filaments; (2) motor concentration, $P_a$, which indicates the fraction of active bonds that are attached by motors and thus induce anti-correlated kicks on the connected node pairs.

By using the independent oscillator approximation which yields sitewise decoupling of the particles, the free energy (due to configurational degrees of freedom) is expressed as a sum of the effective potentials between the interacting density clouds
\begin{equation}
\beta V_\textrm{eff}(|\vec{r}_i-\vec{r}_j^{\,f}|;\alpha_j)=-\ln\int d\vec{r}_j\rho_j(\vec{r}_j)
e^{-\frac{1}{2}\beta U(\vec{r}_i-\vec{r}_j)},
\end{equation}
which essentially averages the Mayer f-bond, $\exp\left[-(1/2)\beta U(\vec{r}_i-\vec{r}_j)\right]$, over the location of the $\vec{r}_j$ particle with an assumed Gaussian density distribution about the fiducial position $\vec{r}_j^{\,f}$
\begin{equation}
\rho(\vec{r}_j, \vec{r}_j^{\,f})=\left(\frac{\alpha_j}{\pi}\right)^{d/2}e^{-\alpha_j(\vec{r}_j-\vec{r}_j^{\,f})^2},
\end{equation}
where $d$ is the spatial dimension of the system.

Self consistency requires that the effective potential mimics the harmonic comparison potential. Thus the curvature of the effective potential must coincide with the phonon frequency or spring constant of the Einstein harmonic oscillators. This gives a coupled set of self-consistent equations for $\{\alpha_i\}$
\begin{equation}
\alpha_i=\frac{1}{2d}\sum_j Tr[\nabla\nabla V_\textrm{eff}(|\vec{r}_{i}^{\,f}-\vec{r}_j^{\,f}|;\alpha_j)].
\end{equation}
In the present work the force constants $\{\alpha_i\}$ will be taken to be equal. This is an effective medium approximation.


For a cat's cradle with excluded volume and subject to correlated motor kicks described by the pair-level interaction (Eqs.~\ref{Teff}--\ref{Ueff}), the effective potential $\widetilde{\beta}_\textrm{eff}\widetilde{V}_\textrm{eff}$ can be expressed as
\begin{equation}\label{SCP_Veff}
e^{-\widetilde{\beta}_\textrm{eff}\widetilde{V}_\textrm{eff}(R,\,\alpha)}=
\sqrt{\frac{\alpha}{\pi}}\frac{1}{R}\int^{\infty}_{0}dw w \left[e^{-\alpha(w-R)^2}-e^{-\alpha(w+R)^2}\right]\times f(w),
\end{equation}
where the function $f(w)$ encodes the pair interaction depending on whether the nodes are bonded and whether the bond is motor-attached. $f(w)$ can be written in terms of the Heaviside step function $\Theta(x)$ as follows
\begin{eqnarray}\label{interaction_kernel}
\nonumber f(w)&=&(1-P_c)\Theta(w-\sigma)\times1\\
\nonumber&&+P_c(1-P_a)\left[\Theta(w-\sigma)\Theta(L_e-w)\times1
+\Theta(w-L_e)\times e^{-\frac{1}{4}\beta\gamma(w-L_e)^2}\right]\\
\nonumber&&+P_cP_a\left[\Theta(w-\sigma)\Theta(L_e-w)\times w^{-\Delta/2}
+\Theta(w-L_e)\times w^{-\Delta/2} e^{-\frac{1}{4}\beta\gamma[1+(s-1/2)\Delta](w-L_e)^2}\right].\\
&&
\end{eqnarray}
Here the first line denotes the interaction between non-bonded pairs where only hard-core repulsion enters ($\sigma$ stands for the hard-core diameter);
the second line is due to the interaction between passively-bonded pairs where elastic bond stretching takes place beyond $L_e$;
the third line accounts for the interaction between actively-bonded pairs where anti-correlated motor kicks induce an effective attraction even in the buckling regime ($\sigma<w<L_e$) and yield an effective bond stiffness that depends on motor activity and susceptibility.

\begin{figure}[htb]
\begin{center}
\centerline{\includegraphics[angle=0, scale=0.55]{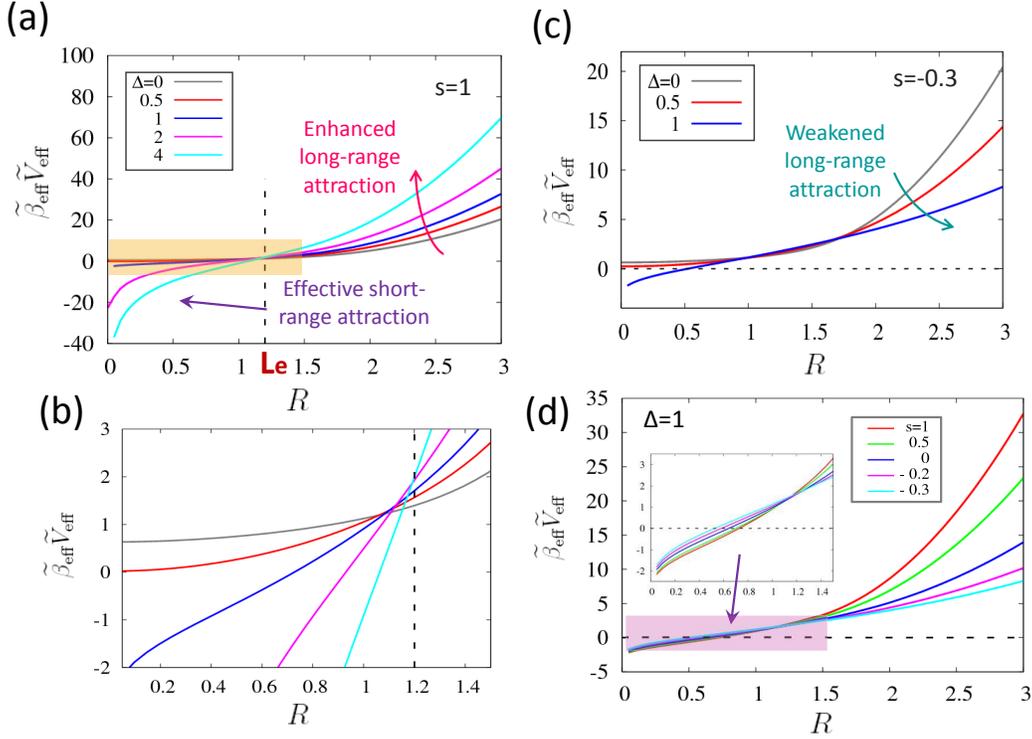}}
\caption{Profile of the effective pair interaction (Eq.~\ref{SCP_Veff}) obtained by the self-consistent phonon calculation.
(a) Susceptible motors ($s=1$) with increasing activity $\Delta$ enhance the long-range attraction and strengthen the short-range effective attraction. (b) A zoom-in view of the small-$R$ region close to the elasticity onset (dashed line) in panel (a) showing the absence of kink or inflection point in the potential profile. (c) Adamant motors ($s=-0.3$) weaken the long-range attraction. No stable $\alpha$ solution is found if the motor activity gets too high ($\Delta>1$). (d) Varying motor susceptibility does not affect the short-range effective attraction (inset), but increasingly susceptible motors (bottom to top) lead to a stronger long-range attraction. Common parameters are $L_e=1.2, \beta\gamma=5, P_c=0.4, P_a=1$.}
\label{Veff_profile}
\end{center}
\end{figure}

Fig.~\ref{Veff_profile} shows the profile of the effective potential $\widetilde{\beta}_\textrm{eff}\widetilde{V}_\textrm{eff}(R,\,\alpha)$ given by Eq.~\ref{SCP_Veff} at various values of motor activity ($\Delta$) and susceptibility ($s$). Panel (a) illustrates the logarithmically growing potential at small separations due to motor-induced effective attraction, and the quadratically increasing potential at large distances due to ordinary bond stretching. Under susceptible motor kicks ($s=1$), both the short-range and long-range attractions become enhanced as the motor activity increases. The dashed line marks the location where the elasticity of individual filaments sets in (i.e. $R=L_e$). A zoom-in view of the small-$R$ region in panel (a) close to the elasticity onset (shaded region) is presented in panel (b). This figure clearly shows that the effective average potential increases monotonically with increasing separation; there is no inflection point or kink in the potential profile which would kinetically slow binding. Thus no difficulty is expected for initial aggregation, as evidenced by simulations. For load-resisting motors (panel c), however, as the motor activity increases, long-range attraction apparently becomes weakened due to a higher $T_\textrm{eff}$, while the susceptibility-independent short-range attraction still gets stronger. No finite $\alpha$ solution can be stabilized when the motor activity gets too high ($\Delta>1$), signaling the development of spatial heterogeneity. Inhomogeneous/Site-dependent $\alpha$ solutions should recover the energy barrier at high motor activity, as seen for the pair-level solution (Fig.3c and d). We demonstrate in panel (d) how the effective potential changes with motor susceptibility. As expected, varying $s$ has little effect on the effective attraction at short distances (more clearly seen in the inset), yet increasingly susceptible motors greatly strengthen the long-range attraction.

In our model, given the asymmetric load response of the nonlinear elastic bonds, the effective attraction between motor-bonded nodes may well drive phase separation of a force-percolating network into dense clumps against voids, i.e. local condensates of contracted bonds (analogous to the droplets of the condensed phase) connected by stretched bonds (associated with surface tension of the droplets), as sketched in Fig.~\ref{tensegrity}.
Spontaneous formation of disordered aggregates has indeed been observed in reconstituted actomyosin systems.\cite{multistage coarsening, active contractility} Computer simulations of the model network when subjected to correlated motor kicks also show that substantial spatial heterogeneity develops when we started from a uniform distribution of motors over an isotropic network (see Fig.~\ref{eff_equil}e for an example).

One thermodynamic indicator of phase separation is a non-monotonic dependence of the pressure upon the density of the constituents. We thus examine how the pressure varies with the node density using the self-consistently determined phonon frequency.
To avoid structural complexity we perform the calculations on regular lattices, yet we expect the same qualitative behavior for random lattices where an isotropic radial distribution function for the fiducial configurations can be obtained from simulations.
The pressure $p$ for a simple cubic lattice with a lattice constant $R$ and a total number of $N$ nodes at the ambient temperature $T$ can be evaluated by numerically differentiating the free energy $F$ as given by
\begin{equation}
p=-\frac{1}{3R^2}\left(\frac{\partial}{\partial R}F\right)_{T,N}k_BT,
\end{equation}
where $F=z\widetilde{\beta}_\textrm{eff}\widetilde{V}_\textrm{eff}(R,\,\alpha;\Delta,s)$ with coordination number $z=6$ for a simple cubic lattice.
The dependence of pressure upon motor properties arises from the effective interaction encoded through $f(w)$ (Eq.~\ref{interaction_kernel}) and from the self-consistent $\alpha$ solution.

[Note that since we have ignored the influence of motor-induced effective interaction on the fiducial structures, the SCP calculation may overestimate the instability threshold of the homogeneous state in terms of motor susceptibility $s$. Nevertheless, given the perturbative nature of the pair solution, any quantitative deviation should be modest.]

\section{Simulations}

To test the validity of the idea of using local effective temperature along with a modified potential for the motorized system, we have performed three types of simulations and compared the resulting steady-state structures. These simulation models are:

(1) Brownian dynamics simulations using a \emph{tensor}-parameter Langevin equation consistent with the effective Fokker-Planck equation (Eq.~\ref{eff FP})

(2) Brownian dynamics simulations at $\bar T_\textrm{eff}$ (Eq.~\ref{Teff}) and with a modified potential $U_\textrm{eff}$ (Eq.~\ref{Ueff}) using a \emph{scalar}-parameter Langevin equation based on the pair-level steady-state solution

(3) Stochastic simulations incorporating anti-correlated kicks along individual active bonds as chemical reaction channels, at bath temperature $T$ and with the bare interaction potential $U(\ris)$. This is of course the most realistic model.



\subsection{Simulation setup}

We investigate a model cytoskeleton that consists of nonlinear elastic filaments subjected to anti-correlated kicks on the motor-bonded node pairs. The \emph{bare} interaction $U(r)$ between bonded nearest-neighbor pairs is taken to be of the cat's cradle type defined earlier, i.e. $\beta U(r)=\Theta(r-L_e)\beta\gamma(r-L_e)^2/2$. The assumed weakness of the excluded volume effect allows large-scale structural rearrangements to occur rather readily.


We build the model network on a simple cubic lattice to avoid structural complexity, and connect the nearest-neighbor nodes with nonlinear elastic bonds at a given probability $P_c$. Note that despite the regular lattice structure, disorder is still inherent in the randomness of bond connectivity for a partially connected network ($P_c<1$). In most of the cases that we will study, we assume $P_a=1$, i.e., all the bonds are motor-attached. The bond connectivity and motor distribution are quenched once initially assigned; there are no bond or motor rupture events. We choose the filament relaxed length $L_e$ to be larger than the lattice spacing (set as the length unit in simulations), so that the initial homogeneous network is completely floppy with no tense bonds at all. The system size is $N=6^3$ and periodic boundary conditions are applied.

Since we are interested in the steady-state behavior at an effective equilibrium, we have chosen a kick step size that is sufficiently small such that higher order contributions to the $l$-expansion are not significant. On the other hand, the kick size is large enough such that the motor-induced effective interaction out-competes the thermal randomization. What affects the dynamics and structural development is the dimensionless motor activity or P\'{e}clet number $\Delta\equiv\kappa l^2/D_0$ which describes the relative strength of the motor kicking noise with respect to the thermal noise. Since in Brownian dynamics formalisms the motor kicking rate $\kappa$ and the kick step size $l$ always appear in combination as $\kappa l^2$, an appropriate kicking rate has been chosen such that $\Delta>1$ yet no instability occurs.
[In stochastic simulations, however, a higher basal kicking rate $\kappa$ yields a faster approach to the steady state without influencing the steady-state features, since the basal kicking rate does not affect the \emph{relative} probability of different kicking events.]

Brownian dynamics (BD) simulations \cite{Brownian dynamics} have been implemented via the position Langevin equation
$\Delta\vec{r}_i(t)=\boldsymbol\mu^i_\textrm{eff}\cdot(-\nabla_iU_\textrm{mod})\Delta t+\vec{R}^\textrm{eff}_i(\Delta t)$,
where $\vec{R}^\textrm{eff}_i(\Delta t)$ represents the random motion due to thermal noise.
The tensor formalism is equivalent to the effective Fokker-Planck equation (Eq.~\ref{eff FP}) yet without hydrodynamic interactions. The total mobility of node $i$ is given by $\boldsymbol\mu^i_\textrm{eff}=\beta(D_0\boldsymbol1+s\kappa l^2\sum_j C_{ij}\hat{r}_{ij}\hat{r}_{ij})$, and the modified interaction force $(-\nabla_iU_\textrm{mod})$ comprises the total mechanical force $(-\nabla_iU)$ acting on node $i$ and the effective interaction $(-\nabla_i\Lambda)$ given by Eq.~\ref{lambda}. The movement due to motor-induced effective attraction thus follows $\boldsymbol\mu^i_\textrm{eff}\cdot(-\nabla_i\Lambda)=\kappa l^2\sum_j C_{ij}\hat{r}_{ij}/r_{ij}$ to quadratic order in $l$.
The scalar formalism, based on the pair-level steady-state solution (Eq.~\ref{pair solution}), sums up the contribution from individual bonded neighbors $j$, i.e., $\boldsymbol\mu^i_\textrm{eff}\cdot(-\nabla_iU_\textrm{mod})\rightarrow\sum_j\boldsymbol\mu^{ij}_\textrm{eff}
\cdot\vec{F}^{ij}_\textrm{mod}=\sum_j\beta(D_0+C_{ij}s\kappa l^2)\vec{F}^{ij}(r_{ij})+\kappa l^2\sum_jC_{ij}\hat{r}_{ij}/r_{ij}$, where $\vec{F}^{ij}(r_{ij})$ is the bare interaction force between node $i$ and its bonded neighbor $j$.
The stochastic dynamics governed by the full master equation (Eq.~\ref{Lne}) has been realized by implementing dynamic Monte Carlo simulations \cite{MC} that obey the model kinetic rate (Eq.~\ref{model rate}).

For making comparisons between the various simulation schemes, we ensure that all runs utilize the same lattice structure, bond connectivity and motor properties. Converging steady-state behavior would then validate the picture of an effective equilibrium at the effective temperature and with the modified interaction as predicted by the expansion.

\subsection{Illustrations}

\subsubsection{Validity of an effective equilibrium and arrested phase separation}

An explicit way to test the validity of picturing the non-equilibrium system driven by small-step motors as being at an effective equilibrium is to compare the steady-state characteristics resulting from the three types of simulations that we described above.

\begin{figure}[htb]
\begin{center}
\centerline{\includegraphics[angle=0, scale=0.55]{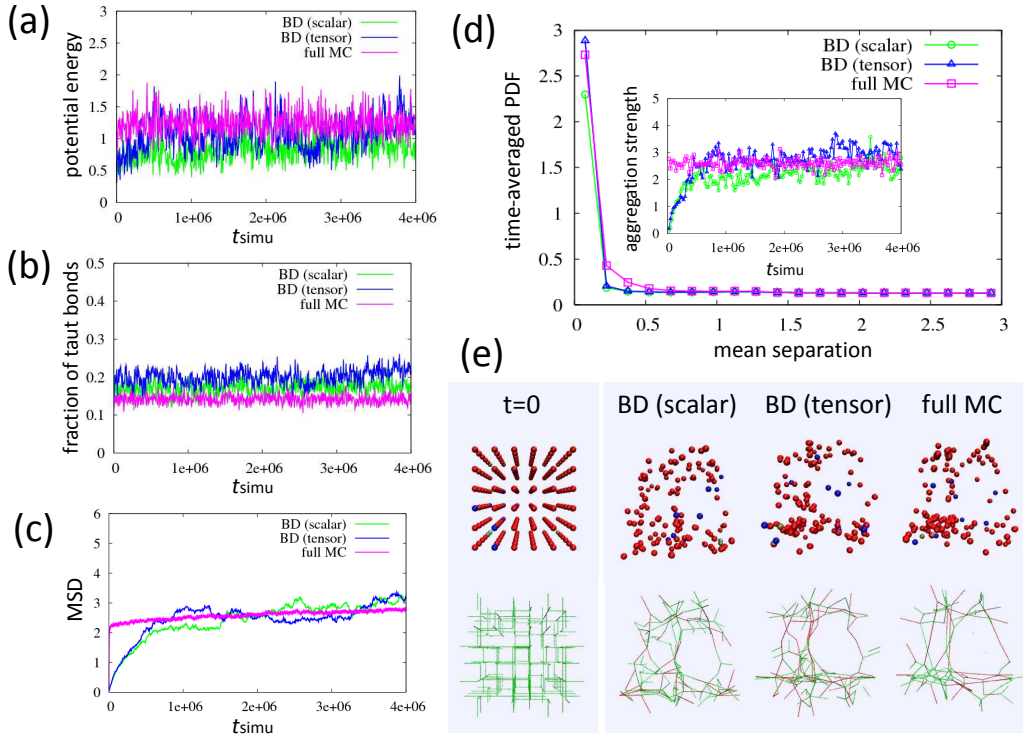}}
\caption{Testing the validity of the effective equilibrium approximation: a comparison of three simulation schemes. Statistical characteristics and steady-state structures for a partially and randomly connected ($P_c=0.4$) network built on a simple cubic lattice driven by small-step ($l=0.03$) susceptible ($s=1$) motors are shown. (a) The potential energy; (b) the fraction of taut bonds; (c) the mean square node displacement; (d) main: the pair distribution function (PDF) averaged over a wide steady-state time window; inset: the aggregation strength, which is the height of the innermost peak of the PDF, versus simulation time; (e) initial (left) and steady-state (right) node configurations (upper row) and corresponding bond structures (lower row). The parameters chosen for illustration are $L_e=1.2, \beta\gamma=5, P_a=1$, and $\kappa=1$.}
\label{eff_equil}
\end{center}
\end{figure}

At a modest kick step size ($l=0.03$), all three simulation schemes lead to quite similar steady-state behavior despite disparate dynamics toward the steady state. The structural characteristics include (1) the mean squared node displacement (MSD) with respect to the initial regular configuration and (2) the amplitude of the innermost peak of the pair distribution function (PDF) which reflects the strength of aggregation.
The PDF is defined as $\textrm{PDF}(r)=C\delta n(r,r+\delta r)/r^2\delta r$ where $\delta n(r,r+\delta r)$ counts the number of particles within an interval $\delta r$ at a distance $r$ from the central particle, and the numeric factor $C$ takes care of normalization.
Both of these measures are almost identical for all the three schemes; they saturate to the same steady-state plateau value (see Fig.~\ref{eff_equil}c, d). The PDF profile in the main panel of Fig.~\ref{eff_equil}d is obtained by averaging over a wide steady-state time window. Note that the compact aggregation, reflected in the steep rise of the PDF in the vicinity of the central node (i.e. the dominant peak at the minimum separation), results from the motor-induced short-range attraction in addition to the absence of excluded volume. The inset shows the time evolution of the aggregation strength. The slight disparity in the potential energy (Fig.~\ref{eff_equil}a) and the fraction of taut bonds (Fig.~\ref{eff_equil}b) seen in each simulation might arise from the perturbative nature of the expansion as well as from the difference in dynamics. As the kick step size increases we would expect larger deviations.
The close resemblance of the steady-state node configurations and bond structures between different schemes (Fig.~\ref{eff_equil}e) lends explicit support to the equivalence of the three schemes in the small kick limit, thus validating the picture of an effective equilibrium with modified interactions.

To demonstrate the arrested phase separation, as anticipated from our theory, the network connectivity has been chosen to be sufficient for tension percolation yet moderate enough to allow considerable local force asymmetry ($P_c=0.4$, i.e. $z=2$--$3$).
The initial network is completely floppy (in green) with no tense bonds (in red) at all and the nodes sit on a simple cubic lattice (Fig.~\ref{eff_equil}e left panel).
As shown in the right panels of Fig.~\ref{eff_equil}e, under anti-correlated susceptible motor kicks, the active nodes (those with motor-attached bonds; shown as red spheres) begin to aggregate and tend to separate from the passive nodes (those with no motor-attached bonds; shown as blue spheres). The corresponding network structure exhibits clumps of floppy bonds (concentrated short green lines) connected by tense bonds (long red lines).
The overall rigidity of the structure is protected by susceptible motors which tune the balance between local bond contraction and neighboring bond stretching such that energetically unfavorable tense states are avoided.

The arrested phase separation builds up as follows:
Susceptible motors enhance the potential gradient via an ($s$-dependent) effective temperature, since $T_\textrm{eff}<T$ along the kicking direction if $s>1/2$. This leads to an enhanced long-range attraction and thus strengthens the initial trend of aggregation among the motor-bonded nodes. The motor-induced short-range attraction then efficiently makes the aggregates become compact, yielding a phase separation of an initially homogeneous structure into node-rich and node-poor regions.
In analogy to the nucleation of liquid drops within an initially homogeneous gas, where large surface tension serves as the driving force to form a bulk condensed phase, in our picture, local aggregates correspond to the droplets of the condensed phase whereas the stretched bonds connecting the aggregates contribute to the surface tension. Therefore, an ensuing coarsening process serves to reduce the surface area via coalescence of local aggregates into larger condensates.
Once a \emph{balance} between local bond contraction and neighboring bond stretching (given a force-percolating network structure) is achieved, the coarsening process stops and the structure does not evolve any further (as reflected in the plateau of the aggregation strength shown in Fig.~\ref{eff_equil}d inset). The system ends up with an arrested heterogeneous structure with compact aggregates/dense clumps coexisting with voids/dilute regions (Fig.~\ref{eff_equil}e right panels); only moderate fluctuations about the arrested structures have been observed.
Therefore, local force asymmetry is necessary for the initiation of phase separation while force percolation is essential for achieving global balance and thus the eventual arrest.

\begin{figure}[htb]
\begin{center}
\centerline{\includegraphics[angle=0, scale=0.5]{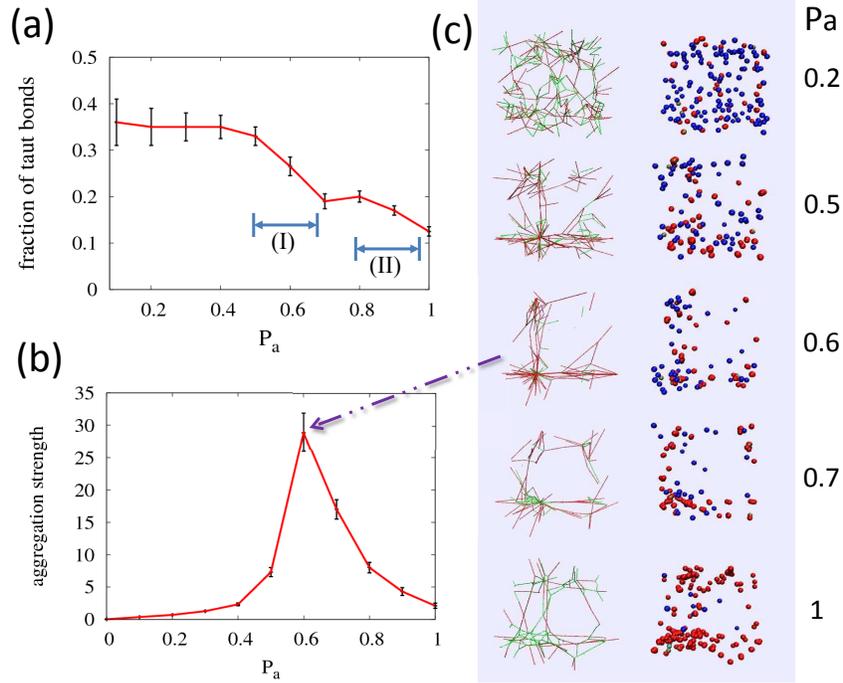}}
\caption{The dependence of network tenseness and structure on motor concentration ($P_a$) obtained by Monte Carlo simulations. The parameters were chosen such that the system is in the regime of arrested phase separation. (a) The fraction of taut bonds decreases as $P_a$ increases. A kink located around $P_a=0.7$ separates two descending branches: (I) $P_a=0.5$--$0.7$ and (II) $P_a=0.8$--$1$. (b) The aggregation strength exhibits a sharp peak at $P_a=0.7$.
The error bars in (a) and (b) depict standard deviations from averages over a steady-state time window of $4\times10^6$ Monte Carlo steps.
(c) Bond structures and corresponding node configurations at various $P_a$ values are shown, from top to bottom $P_a=0.2, 0.5, 0.6, 0.7$ and $1$. The arrow indicates the bond structure with the strongest aggregation. The remaining simulation parameters are $L_e=1.2, \beta\gamma=5, P_c=0.4, l=0.05, s=1$, and $\kappa=0.1$.}
\label{ftb_Pa}
\end{center}
\end{figure}

In Fig.~\ref{ftb_Pa} we demonstrate how the concentration $P_a$ (and thus the cooperativity) of susceptible motors ($s=1$) affects the tenseness and structural organization of the network. Parameters are chosen such that the system is in the regime of arrested phase separation at sufficiently high $P_a$. Shown in panel (a), the overall trend of a decrease in the fraction of taut bonds as $P_a$ is raised is apparent. This results from increasing cooperativity in local force generation. Particularly noteworthy, however, is the presence of a kink near $P_a=0.7$, which separates two descending branches (I: $P_a=0.5$--$0.7$ and II: $P_a=0.8$--$1$ as marked in the figure). In stark contrast, the aggregation strength (i.e. the amplitude of the first peak of the radial distribution function) exhibits a highly \textit{non-monotonic} dependence on $P_a$, which is sharply peaked at $P_a=0.6$ (see panel b). The total energy (not shown) shows a similar trend as $P_a$ varies.
To understand these trends, we visualize the steady-state bond structures and corresponding node configurations in panel (c). These simulated configurations reveal two distinct regimes characterized by different ``strategies" to reduce the number of taut bonds:
At intermediate motor concentrations (corresponding to branch I), as $P_a$ rises, an increasing number of floppy bonds are formed at the aggregation centers, at the cost of fewer but even more strongly stretched inter-clump filaments due to more compact node aggregation, as most clearly seen for $P_a=0.6$ in panel (c), where the corresponding aggregation strength of nodes reaches its peak value (indicated by arrow).
At sufficiently high motor concentrations (branch II), however, cooperative motor processes tend to minimize the overall tenseness by buckling as many bonds as possible at the same time avoiding too strong stretching. The resulting structure is thus rich in \textit{moderately buckled} bonds and consists of large floppy clumps connected by sparse tense bonds (see $P_a=1$ case in panel c). The aggregation strength decreases with increasing $P_a$ in this highly cooperative regime.
Such an interesting dependence of structural development on motor concentration vividly demonstrates the intricate interplay of local force generation and collective motor action.

\subsubsection{Effective attraction}

\begin{figure}[htb]
\begin{center}
\centerline{\includegraphics[angle=0, scale=0.6]{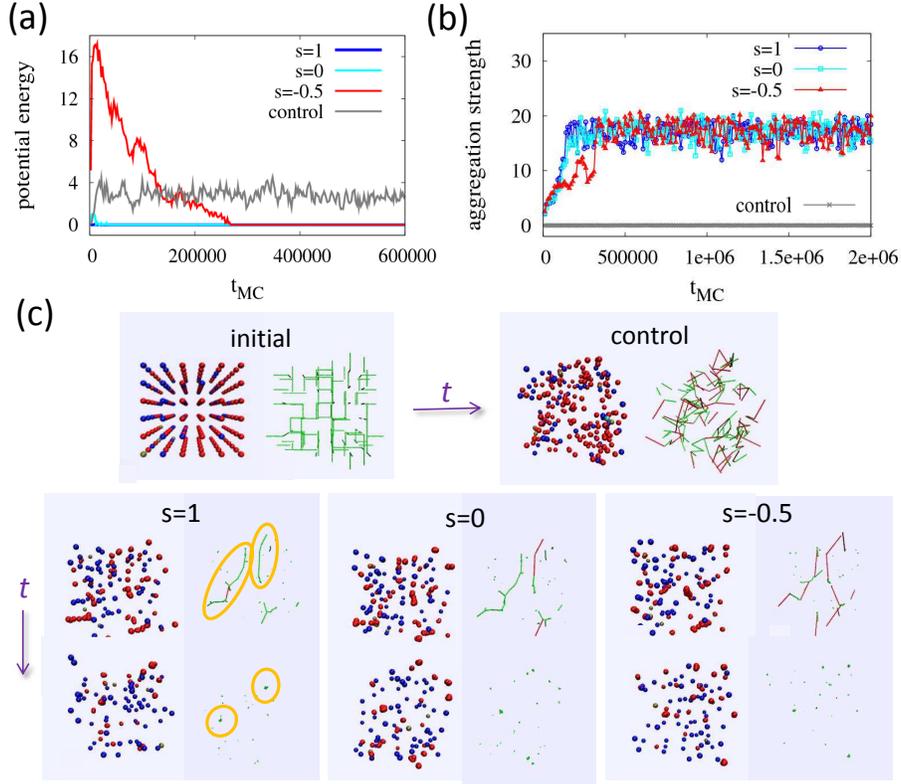}}
\caption{An illustration of the motor-induced effective attraction for a non-percolating network ($P_c=0.2$) at various motor susceptibilities. (a) The potential energy; (b) the aggregation strength; (c) initial (upper left) and later node configurations and bond structures for a control run with pure thermal motion (upper right), and for motorized systems with $s=1$ (lower left), $s=0$ (lower middle) and $s=-0.5$ (lower right). Despite having different dynamics, similar steady-state structures with isolated floppy clumps are reached in each case, regardless of the motor susceptibility. The remaining simulation parameters are $L_e=1.2, \beta\gamma=5, P_a=1, l=0.03$, and $\kappa=0.1$.}
\label{eff_attraction}
\end{center}
\end{figure}

To demonstrate the effect of motor-induced short-range attraction, we choose a network connectivity that is below the percolation threshold ($P_c=0.2$).
In the absence of global force percolation, when driven by spatially anti-correlated motors kicks, the initially homogeneous and entirely floppy network (Fig.~\ref{eff_attraction}c upper left panel) develops into isolated floppy clumps (Fig.~\ref{eff_attraction}c lower panels), regardless of motor susceptibility. Under susceptible motor kicks ($s=1$), it becomes evident that further contractions of the already buckled bonds (with typical spots marked by circles in Fig.~\ref{eff_attraction}c lower left panels) arise solely from the effective short-range attraction due to motor processes. On the other hand when driven by load-resisting motors ($s=-0.5$), the anti-correlation in movements causes collapse of the transiently stretched bonds (Fig.~\ref{eff_attraction}c lower right panels).
Adamant motor kicks ($s=0$) lead to a similar transient structure (with fewer tense bonds though) and the eventual collapse (Fig.~\ref{eff_attraction}c lower middle panels).
Therefore in a non-percolating network driven by anti-correlated kicks, despite the very different dynamics due to differing motor susceptibility, similar steady state structures are reached.
The complete collapse of all the individual clumps is characterized by a vanishing total potential energy after the initial transients (Fig.~\ref{eff_attraction}a) and a significant aggregation strength that saturates to a steady-state plateau (Fig.~\ref{eff_attraction}b) once isolated condensates form.

In contrast, the control run with pure thermal motion presents a considerable fraction of taut bonds (Fig.~\ref{eff_attraction}c upper right panel) and thus maintains a finite potential energy (grey curve in Fig.~\ref{eff_attraction}a). The bond structure and node configuration remain largely homogeneous, exhibiting modest density fluctuations and a vanishingly small aggregation strength (grey curve in Fig.~\ref{eff_attraction}b).



Note that the illustrations for effective attraction shown here are obtained using fully stochastic simulations. The effective Brownian dynamics schemes give similar steady state structures composed of isolated floppy clumps only for susceptible motor kicking. For load-resisting motors that give rise to an effective long-range repulsion, however, a distinct behavior is seen (detailed below), highlighting the significance of correlation in motion for structural development.

\subsubsection{Effective repulsion}

\begin{figure}[htb]
\begin{center}
\centerline{\includegraphics[angle=0, scale=0.46]{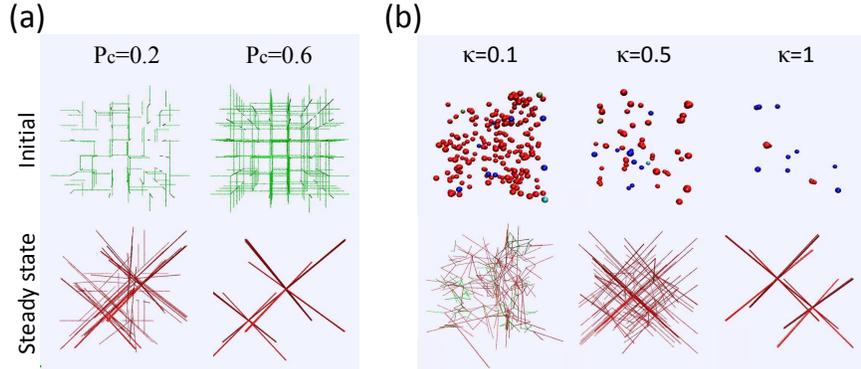}}
\caption{An illustration of the motor-induced effective repulsion caused by load-resisting ($s=-0.5$) motors for various network connectivities and motor kicking rates. (a) Initial (upper) and steady-state (lower) bond structures are shown at low (left: $P_c=0.2$) and at high (right: $P_c=0.6$) connectivity with $\kappa=1$. (b) Node configurations (upper) and corresponding bond structures (lower) at various motor kicking rates (left to right: $\kappa=0.1, 0.5$, and $1$) with $P_c=0.4$. The common set of simulation parameters are given by $L_e=1.2, \beta\gamma=5, P_a=1$, and $l=0.05$.}
\label{eff_repulsion}
\end{center}
\end{figure}

Another interesting case arises when the motor susceptibility becomes negative. In this case a negative effective temperature yields an effective repulsion at sufficiently high motor activity (refer to Eqs.~\ref{Teff}--\ref{Ueff}).
As shown in Fig.~\ref{eff_repulsion}(a), starting with an entirely floppy network (upper row), enhanced bond stretching coming from thermally induced fluctuations in bond length giving local force asymmetries, eventually gives rise to highly tense and ordered aster patterns (lower row) at steady state. Aster formation occurs when the effective repulsion that promotes node separation and thus bond stretching dominates over the effective attraction that drives the opposite; removal of the short-range effective attraction does not affect aster formation, but indeed does disrupt efficient aggregation (see Fig.~\ref{aggregation_kappa}e). Consistent with our earlier results for \textit{uncorrelated} kicks,\cite{spontaneous motion} the effective Brownian dynamics simulations also give sustained aster patterns. These patterns cannot collapse due to the absence of pairwise anti-correlation in motion; since at each move in Brownian dynamics, an individual node sees only an effective potential due to \textit{all} its neighbors, the pairwise correlation is virtually lost. Such pairwise anti-correlation is crucial for active contractility as we showed elsewhere.\cite{actomyosin contractility}
This anti-correlation, however, is captured by complete Monte Carlo simulations where anti-correlated move pairs are treated as reaction channels and executed at each MC step.

Fig.~\ref{eff_repulsion} also highlights the architectural and dynamical ingredients required for the formation of connected tight asters that consist of tense bundles radiating from the common center. Panel (a) illustrates the necessity of a sufficient network connectivity for force transmission and bundle compaction; at low connectivity (Fig.~\ref{eff_repulsion}a left) only individually separate tense bundles are formed.
Panel (b) depicts that a high motor kicking rate is needed to defeat thermal spreading and thus to facilitate filament or bundle alignment.

\begin{figure}[htb]
\begin{center}
\centerline{\includegraphics[angle=0, scale=0.35]{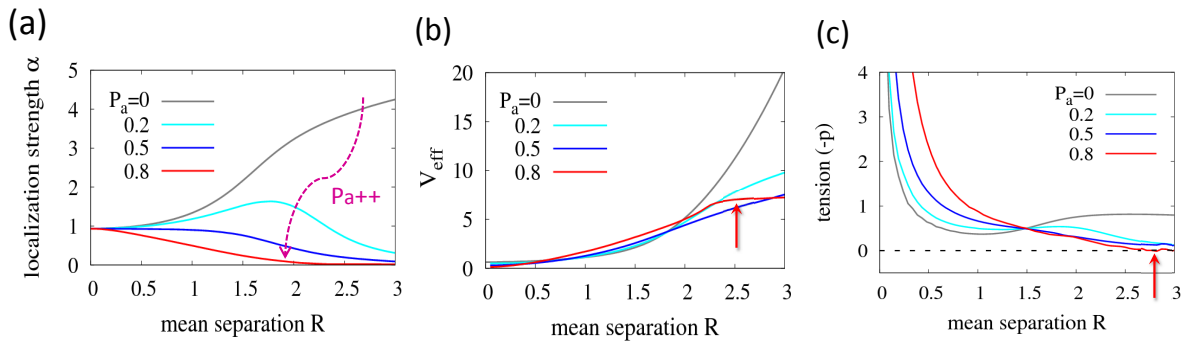}}
\caption{Mean-field predictions of the effect of the concentration ($P_a$) of load-resisting ($s=-0.5$) motors on long-range interactions. (a) The localization strength $\alpha$ of individual nodes. The localization at large separation $R$ is considerably suppressed as $P_a$ increases.
(b) The effective potential. Increasing $P_a$ weakens the long-range attraction; the potential profile actually flattens out (red arrow) at $P_a=0.8$ indicating a vanishing restoring force. (c) The overall tension $(-p)$ vanishes at large $R$ (red arrow) for high $P_a$. This suggests the tendency for contraction is counterbalanced by a motor-induced long-range repulsion.
The simulation parameters are $L_e=1.2, \beta\gamma=5, P_c=0.4$ and $\Delta=1$.}
\label{vary_Pa}
\end{center}
\end{figure}

Aster formation finds a natural explanation
in our model when we use the notion of the effective long-range repulsion that we have derived.
In Fig.~\ref{vary_Pa} we show the mean-field indications, obtained by SCP calculations (described in section II.C), of how the concentration ($P_a$) of load-resisting ($s=-0.5$) motors affects the long-range interactions.
As clearly seen in panel (b), as $P_a$ increases the long-range attraction due to bond stretching considerably weakens. Accordingly, the localization strength $\alpha$ of individual nodes (panel a) and the tension $(-p)$ within the network (panel c) are both strongly suppressed. At $P_a=0.8$, the profile of the effective potential becomes almost flat at large distances (red arrow in panel b), indicating a vanishing restoring force. Consistently, $\alpha$ becomes vanishingly small at large $R$ (panel a) and the overall tension decays to zero (red arrow in panel c), because the bond constraints are hardly felt when load-resisting motors counteract the tendency to contract. Yet higher motor concentration still enhances the effective short-range attraction (small-$R$ region in panel c) as expected.




\subsubsection{Effect of motor activity and susceptibility in phase separation: mean-field indications}

By performing the self-consistent phonon calculation, we find mean-field indications for the phase separation observed in the simulations and experiments. In particular, such calculation allows us to examine how the changes in motor activity affect the tendency to phase separate which can be tested against simulations.
We choose the length unit to be $100\sigma$ (such that the excluded volume effect plays a negligible role in phase separation) and vary the mean separation $R$ between the nearest neighbors, i.e. the lattice spacing of the simple cubic lattice.

\begin{figure}[htb]
\begin{center}
\centerline{\includegraphics[angle=0, scale=0.52]{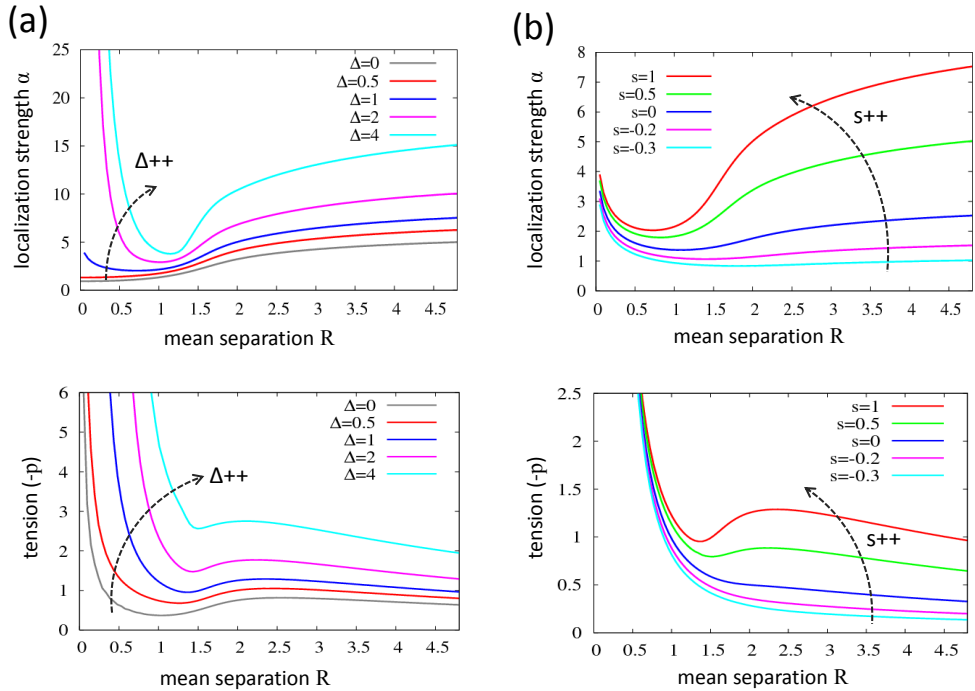}}
\caption{The effect of motor activity ($\Delta$) and susceptibility ($s$) on phase separation. Shown are the calculated localization strength $\alpha$ (upper row) and the tension ($-p$) (bottom row) as a function of the mean separation $R$ for (a) $s=1$ with various motor activities, $\Delta=0, 0.5, 1, 2$, and $4$ (bottom to top), and (b) $\Delta=1$ with various motor susceptibilities, $s=-0.3, -0.2, 0, 0.5$, and $1$ (bottom to top). The remaining simulation parameters are $L_e=1.2, \beta\gamma=5, P_c=0.4$, and $P_a=1$.}
\label{SCP_motor_effect}
\end{center}
\end{figure}

At sufficiently high motor susceptibility ($s>0$ for $\Delta=1$; Fig.~\ref{SCP_motor_effect}b lower panel), we observe a non-monotonic dependence of the tension ($-p$), i.e. negative pressure, on the mean separation $R$, clearly indicating the necessity of some kind of phase separation into node-rich and node-poor regions.

For susceptible motors with $s=1$ (Fig.~\ref{SCP_motor_effect}a), varying the motor activity $\Delta$ affects \emph{both} the short-range (small $R$) and the long-range (large $R$) attractions: increasing motor activity (as indicated by dashed arrows) leads to stronger localization of individual nodes (upper panel) and a larger tension in favor of stronger aggregation (lower panel). In particular at small $R$ where the bonds are buckled, effective attraction arising from motor-driven contractions ($\sim\Delta\log r$) dominates, yielding further aggregation of the loosely connected nodes. This behavior is most clearly manifested for a sparsely connected network where lack in bond constraints allows the formation of isolated aggregates, as shown earlier in Fig.~\ref{eff_attraction}c.

\begin{figure}[htb]
\begin{center}
\centerline{\includegraphics[angle=0, scale=0.58]{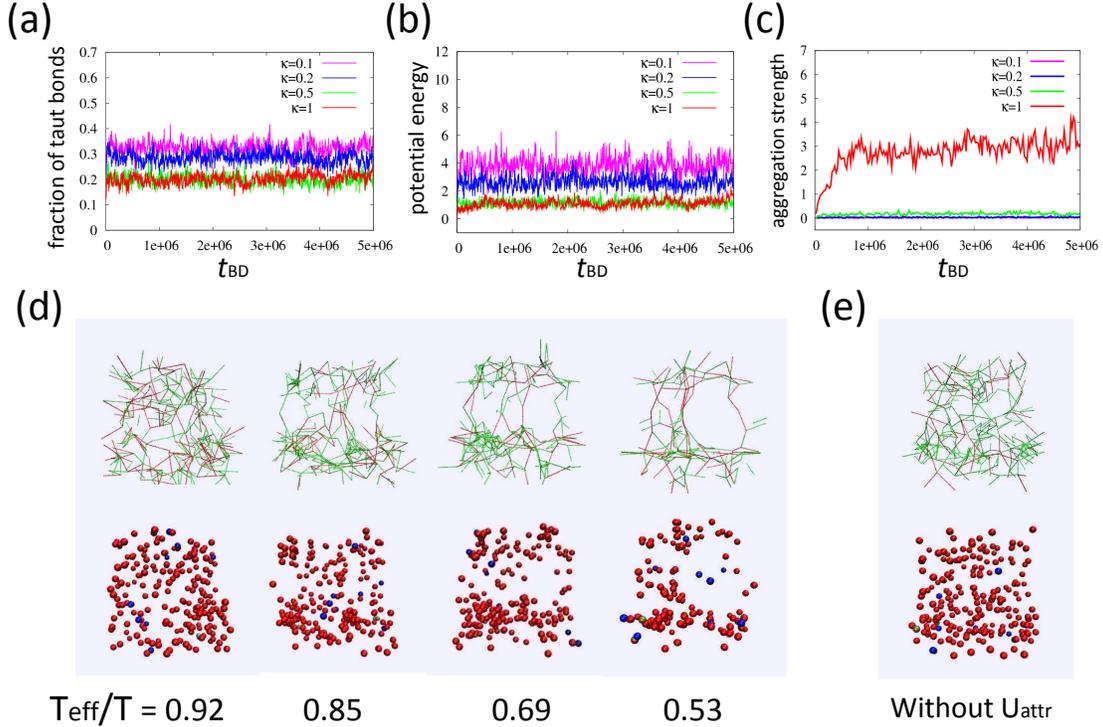}}
\caption{The role of motor kicking rate and effective attraction in aggregation. (a)--(c) Statistical measures for the dynamic and structural development at various motor kicking rates. Steady-state bond structures (upper) and node configurations (lower) are shown at increasing motor kicking rates (d): from left to right $\kappa=0.1, 0.2, 0.5$ and $1$ (converted into $T_\textrm{eff}/T$), and for corresponding models without motor-induced short-range attraction at $\kappa=1$ (e). The remaining simulation parameters are $L_e=1.2, \beta\gamma=5, P_c=0.4, P_a=1, s=1$, and $l=0.03$.}
\label{aggregation_kappa}
\end{center}
\end{figure}

Using Brownian dynamics simulations,
we study the dependence of aggregation strength upon motor activity for force-percolating networks ($P_c=0.4$). The statistical measures and steady-state structures (labeled by $T_\textrm{eff}/T$) are displayed in Fig.~\ref{aggregation_kappa}. Increasing motor kicking rate $\kappa$ (note $\Delta\propto\kappa$) apparently enhances
the trend toward phase separation (panel d) and promotes stronger aggregation (panel c and d), supporting the mean-field prediction. Both the fraction of taut bonds (panel a) and the total energy (panel b) decrease with increasing kicking rate due to a lower $T_\textrm{eff}$.

As we pointed out earlier in deriving the effective pair potential, varying motor susceptibility $s$ affects \emph{only} the long-range interaction (via $\boldsymbol{T_\textrm{eff}}$), as shown in Fig.~\ref{SCP_motor_effect}b. For a given lattice spacing, increasingly susceptible motors (indicated by dashed arrows) drive stronger attraction (lower panel) and enhance localization of individual nodes (upper panel), as well as yield a stronger tendency for phase separation, as reflected in the increasingly non-monotonic dependence of the tension on density change as $s$ increases (lower panel).

\begin{figure}[htb]
\begin{center}
\centerline{\includegraphics[angle=0, scale=0.35]{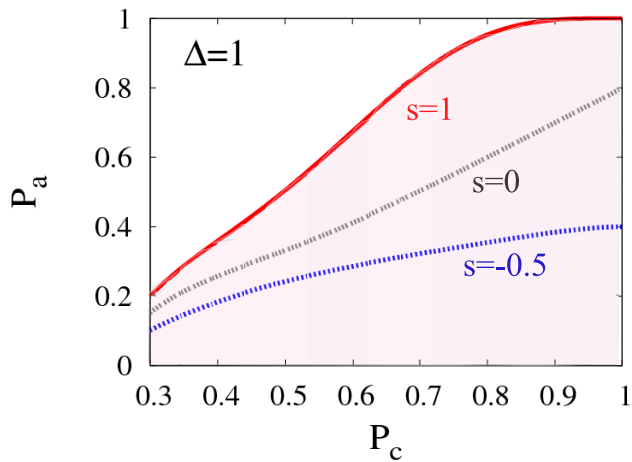}}
\caption{The stability diagram at various motor susceptibilities. The colored lines represent the stability boundaries, solid red for $s=1$, dashed grey for $s=0$ and dotted blue for $s=-0.5$.
Below the stability boundaries the pressure exhibits a non-monotonic dependence on particle separation indicating the tendency toward phase separation. The instability region (shaded area) extends to lower $P_c$ and higher $P_a$ as $s$ increases, suggesting that susceptible motors promote phase separation.
The remaining simulation parameters are $L_e=1.2, \beta\gamma=5,$ and $\Delta=1$.}
\label{stability_diagram}
\end{center}
\end{figure}

Self-consistent phonon calculations also allow us to determine a quasi-thermodynamic stability diagram. In Fig.~\ref{stability_diagram} we present the stability diagram in the $P_c$-$P_a$ parameter plane for susceptible ($s=1$), adamant ($s=0$) and load-resisting ($s=-0.5$) motors. Below the stability boundaries (colored lines), the pressure depends non-monotonically on the mean particle separation. This indicates the tendency toward phase separation. Above the boundaries, there are no stable $\alpha$ solutions , or that the tension (or negative pressure) decreases monotonically with increasing particle separation. As the motor susceptibility increases, the instability region (shaded area) expands toward lower $P_c$ and higher $P_a$, suggesting that susceptible motors promote phase separation. Since motor susceptibility affects only the long-range interaction via $T_\textrm{eff}$ at high $P_c$, where bond stretching stabilizes finite $\alpha$ solutions, small or negative $s$ may lead to an effective repulsion that counteracts the trend of attraction thus eliminating the non-monotonicity in pressure, or else destabilizes $\alpha$ solutions by offsetting the restoring force. Consequently, the stability boundary at large $P_c$ shifts toward lower $P_a$ as $s$ decreases.

To determine whether the non-monotonicity in pressure indeed corresponds to phase separation, we need to examine heterogeneous/site-dependent $\alpha$ solutions; a bimodal distribution of stable $\alpha$ values would then indicate that localized dense regions phase separate from mobile dilute regions.
We hope to investigate this aspect in an upcoming work and thus provide more quantitative arguments for the surface tension associated with stretched bonds connecting floppy clumps. Such an analysis should allow us to determine the ``droplet" size for the condensed phase when bond-constraint-induced arrest occurs.

\begin{figure}[h]
\begin{center}
\centerline{\includegraphics[angle=0, scale=0.35]{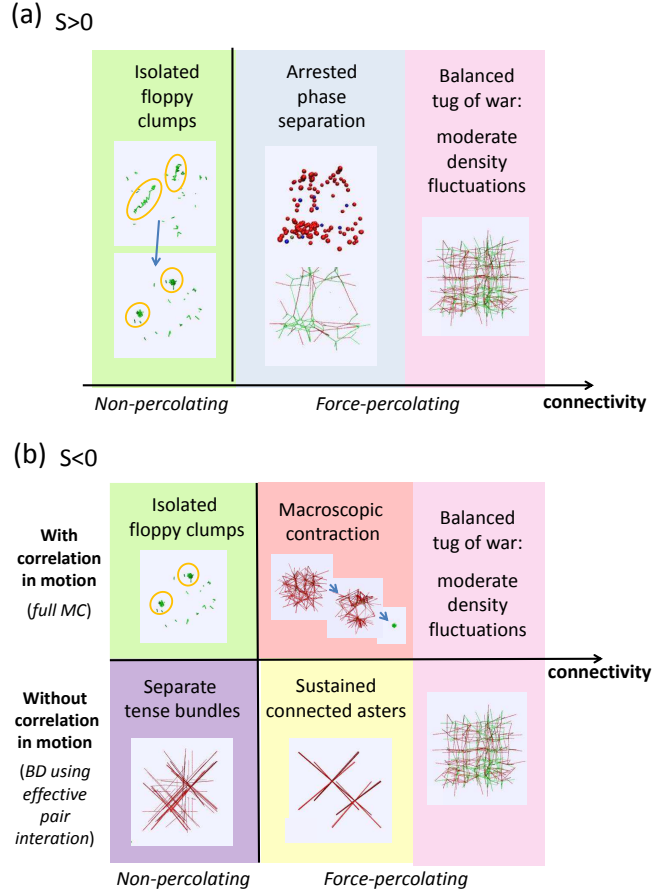}}
\caption{Patterns of behavior for susceptible ($s>0$) and load-resisting ($s<0$) motors at a high motor concentration. Typical structures generated by simulations are shown for each situation.
The horizontal axis indicates increasing network connectivity from left to right. The vertical line locates the percolation threshold.
(a) For susceptible motors, effective Brownian dynamics simulations and Monte Carlo simulations give similar results. At intermediate connectivity above the percolation threshold, arrested phase separation occurs. (b) For load-resisting motors,
(anti-)correlation plays a key role in the active patterning. Macroscopic contraction occurs only in the presence of anti-correlation in motion, otherwise connected asters form that cannot collapse.}
\label{pattern_diagram}
\end{center}
\end{figure}

To summarize our results in Fig.~\ref{pattern_diagram} we outline the diverse patterns formed at various network connectivity ($P_c$) for susceptible ($s>0$) and load-resisting ($s<0$) motors.
This figure delineates where there are contrasting results from complete MC simulations to those obtained from BD simulations using the effective pair interaction.
As we mentioned before, the Monte Carlo schemes explicitly incorporate pairwise anti-correlation by treating move pairs as reaction channels, while in the effective Brownian dynamics schemes the total effective potential acting on individual nodes from all the neighbors smears out the pair correlation.
This disparity between the predictions of the two schemes does not affect the active patterning by susceptible motors (Fig.~\ref{pattern_diagram}a). In that case the effective attraction provides the dominant mechanism for aggregation and phase separation. MC and BD simulations give converging results: Below the percolation threshold (marked by the vertical line) isolated floppy clumps form, whereas at intermediate $P_c$ above the threshold arrested phase separation occurs.
For load-resisting motors (Fig.~\ref{pattern_diagram}b), however, correlation in motion plays a key role in structural development, because in this case an effective long-range repulsion dominates over the short-range attraction and governs the pattern formation.
Anti-correlation in movements gives rise to collapse of the tense intermediates which is essential for active contractility/macroscopic contraction.\cite{actomyosin contractility} If there is no anti-correlation collapse does not occur. As a result, the tense bundles (for non-percolating case) and connected asters (for percolating case) are maintained as long-time steady state structures.
At sufficiently high connectivity, the bond constraints are too strong to allow significant local force asymmetry, thus a balanced tug-of-war between motor-attached filaments prevents the formation of heterogeneous cluster structures, and instead leads to a largely homogeneous structure with moderate fluctuations (rightmost regime in both panels a and b), regardless of the motor susceptibility or correlation in motion.

Also we note that the Brownian dynamics simulations for load-resisting motors exhibit similar behavior to that coming from uncorrelated isotropic kicks as we studied earlier \cite{spontaneous motion}: Both simulations generate sustained aster patterns that do not collapse, exemplifying a negative effective temperature.
For susceptible motors, however, the dependence of $T_\textrm{eff}$ and the effective attraction on the instantaneous local network structure hinders the system from achieving global concerted movement. There is thus no spontaneous flow or oscillating behavior that presents for the uncorrelated kick case.\cite{spontaneous motion} A similar absence of a flow transition has also been found for a contractile nematic model recently studied analytically \cite{viscosity divergence} and numerically.\cite{shear active gels}




\section{conclusion and discussion}

We are now in a position to recapitulate how the intricate interplay between local force generation, network connectivity and collective action of motors gives rise to regular and heterogeneous patterns, arrested coarsening and macroscopic contraction:
A sufficient connectivity is required for forces to percolate so that local motor-induced stresses and resulting deformations can propagate significant distances through the network. Given a force-percolating structure, downhill-prone motors yield heterogeneous/disordered cluster structures, exemplified as an arrested phase separation in the absence of bond or motor rupture events; whereas load-resisting motors may drive large-scale contraction by surmounting a high energy barrier constituted by tense intermediates. Anti-correlation in movements is essential for collapse of the intermediate tense states in approach to the eventual large-scale contraction; in the absence of correlation in movements, as is the case for Brownian dynamics simulations and for our earlier studies on uncorrelated kicks,\cite{spontaneous motion} the stretched bundles cannot collapse and no contractile structures result, thus the aster pattern is maintained as the steady-state structure.

The notion of effective interaction provides a natural explanation for the aggregation and coalescence of actomyosin condensates: enhanced long-range attraction facilitates initial density fluctuations; effective attraction at short distances especially in the buckling regime, arising purely from motor-driven contractions, then promotes efficient aggregation by drawing nearby nodes or condensates even closer. Whereas an effective repulsion can originate from a negative motor susceptibility, and in turn a negative effective temperature allows the formation of aster patterns in the absence of correlation in node movements.

Our finding may suggest a new mechanism for aggregation of active gels: local force asymmetry and disorder (structurally inherent or thermally generated) trigger local aggregation which is further enhanced by an effective attraction due to correlated motor kicks; force percolation combined with the tendency to reduce surface tension associated with the stretched bonds leads to coarsening of local aggregates; when the balance between local bond collapse and neighboring bond stretching is reached, the system forms an arrested structure composed of floppy clumps connected by tense bonds. The pertinent dynamic process involves phase separation into node-rich and node-poor regions followed by arrest due to bond constraints.

We also generalize the concept of effective temperature to non-equilibrium many-body systems driven by \emph{correlated} small-step motor kicking events.
Explicit simulation tests lend support to the validity of picturing such systems as being at an effective equilibrium with modified interactions.
\\











Support from the Center for Theoretical Biological Physics sponsored by the National Science Foundation (Grant PHY-0822283) is gratefully acknowledged.



\end{document}